# Fractional High-Chern Insulator in Twisted Rhombohedral Graphene

Zexu Li[1†], Wenxuan Wang[1†*], Fajie Wang[1†], Zaizhe Zhang[1], Qiu Yang[1,2], Kenji Watanabe[3], Takashi Taniguchi[4], X.C. Xie[1,5,6], Jie Wang[1,7], Kaihui Liu[2*], Zhida Song[1,6,7*] and Xiaobo Lu[1,7*]

[1]International Center for Quantum Materials, School of Physics, Peking University, Beijing 100871, China
[2]State Key Laboratory for Mesoscopic Physics, Frontiers Science Centre for Nano-optoelectronics, School of Physics, Peking University, Beijing 100871, China
[3]Research Center for Electronic and Optical Materials, National Institute of Material Sciences, 1-1 Namiki, Tsukuba 305-0044, Japan
[4]Research Center for Materials Nanoarchitectonics, National Institute of Material Sciences, 1-1 Namiki, Tsukuba 305-0044, Japan
[5]Interdisciplinary Center for Theoretical Physics and Information Sciences, Fudan University, Shanghai 200433, China
[6]Hefei National Laboratory, Hefei 230088, China
[7]Collaborative Innovation Center of Quantum Matter, Beijing 100871, China

[†]Z.L., W.W. and F. W. contributed equally to this work.
*E-mail: xiaobolu@pku.edu.cn; songzd@pku.edu.cn; khliu@pku.edu.cn; wenxuanwang@pku.edu.cn

**The realization of fractional Chern insulators opens up the possibility of exploring fractionally charged excitations[1–7] and anyonic statistics[8–11] in the absence of a magnetic field. A central question is whether lattice-based systems can give rise to radically new states, distinct from those observed in traditional fractional quantum Hall systems[12–20]. In this work, we investigate a new type of moiré flat band system composed of Bernal bilayer graphene and rhombohedral tetralayer graphene. We discover an unprecedented richness of quantum anomalous Hall insulators with Chern numbers from $|C| = 1$ to $|C| = 7$ at $\nu = 1$ and around $\nu = 3$. Remarkably, we observe an exotic fractional Chern insulator with $C = 7/3$ around $\nu = 2/3$ which is beyond all known fractional Chern insulators described by either the Jain sequence or current high Chern theory[21–27]. Our work expands the understanding of fractionally charged excitations beyond the Landau level basis and offers a new moiré platform for exploring anyons.**

Topological phases driven by strong interactions remain a central focus in condensed-matter physics, offering routes to fractionalized excitations and nontrivial quantum order beyond conventional band descriptions. Fractional Chern insulators (FCIs) represent a particularly compelling class of such phases: lattice analogues of fractional quantum Hall states that emerge in topological bands without net magnetic flux[1–7]. Their realization relies on a delicate interplay between band geometry and interactions, making them a powerful platform for exploring correlation-induced topology in engineered quantum materials[8–11,28,29]. While intrinsic FCIs have been experimentally observed in different flat-band systems including twisted $MoTe_2$[12–15] and rhombohedral graphene multilayers[16–20,30–32], most experimental efforts have focused on bands with Chern number $C = 1$. In contrast with the quantum Hall system, a Chern insulator can host higher Chern numbers, without introducing additional degrees of freedom. Theoretical work suggests that bands with higher Chern number can stabilize richer fractional excitations with Abelian states and non-Abelian orders[21–27]. However, achieving such states in experiment has remained elusive. High Chern bands typically feature more complex orbital textures and pronounced Berry-curvature variations that challenge the formation of robust fractional phases, and clear signatures of topological order—such as quantized response and characteristic edge structure—have not yet been established. Rhombohedral graphene has recently emerged as an exceptionally promising platform for realizing FCIs[16–20,30–33] and higher Chern band[33–39]. Owing to its stacking order, rhombohedral graphene naturally hosts flat bands with nontrivial Berry curvature, providing the essential ingredients for stabilizing interaction-driven topological states.

In this work we introduce a new moiré system composed of Bernal bilayer graphene and rhombohedral tetralayer graphene. As shown in Fig. 1a, Bernal bilayer graphene and rhombohedral tetralayer graphene with twist angle $\theta$ are sandwiched between two graphite gates with hBN dielectric layers, forming a moiré superlattice with lattice period $\lambda_m$. The graphite gates enable us to independently control moiré filling $\nu$ and displacement field $D/\varepsilon_0$. Through a high-throughput fabrication method (Method), we have fabricated a series of twisted bilayer rhombohedral tetralayer graphene (TBRTG) devices with twist angle $\theta$ ranging from 1.28° to 1.38°(Extended Data Table. 1). Our subsequent results mainly focus on device D1 with $\theta$ = 1.38°. The non-interacting band structure is shown in Fig. 1c-d with projection illustrated in Fig. 1b. The parameter $\Delta_U$ denotes the electrostatic potential energy difference between adjacent graphene layers, defined as the energy change of an electron moving from the upper layer to the lower layer. In the devices, $\Delta_U$ can be tuned continuously by the displacement field $D/\varepsilon_0$. As $\Delta_U$ increases, the band undergoes a topological transition from $C$ = 4 to $C$ = 3. Figure 1f and g display the longitudinal resistivity $\rho_{xx}$ and Hall resistivity $\rho_{xy}$ as a function of moiré filling factor $\nu$ and displacement field $D/\varepsilon_0$ in device D1 respectively. The filling factor $\nu$ is carefully calibrated by magneto transport measurement (Method).

The phase diagram is asymmetric under positive and negative $D$ fields, consistent with absence of mirror symmetry in the system (Supplementary Fig. 3-4). Moreover, rich correlation phenomena within the flat bands are observed. For instance, a correlated insulator similar to magic-angle graphene emerges at $\nu$ = 2[40,41]. Interestingly, we observe even richer features near $\nu$ = 1 and $\nu$ = 3. In these regions, the longitudinal resistivity $\rho_{xx}$ shows minimums or even vanishes, while the Hall resistivity $\rho_{xy}$ exhibits maximums, suggesting the presence of topological states.

## *D* field modulated Chern insulators

Figures 2a-b display $\rho_{xx}$ and $\rho_{xy}$ near $\nu$ = 1 measured in magnetic fields of $B_\perp$ = ± 0.2 T which are symmetrized or anti-symmetrized accordingly (Method). In most displayed $D$ fields, $\rho_{xx}$ tends to zero, and $\rho_{xy}$ remains constant at $\nu$ = 1 with varying $D$ field. When $D/\varepsilon_0$ > 0.75 V/nm, a new state emerges, showing minimal residual $\rho_{xx}$ and larger $\rho_{xy}$. Figure 2c presents a vertical linecut near $\nu$ = 1 taken from Figure 2a. Notably, this state at $\nu$ = 1 can be precisely quantized to $\rho_{xy}$ = $h/4e^2$ and $h/3e^2$ under different $D$ fields. Figures 2f-g further show $\rho_{xx}$ and $\rho_{xy}$ from the linecuts marked by black and white dashed lines in Fig. 2a, respectively. Both states exhibit vanishing $\rho_{xx}$ with $\rho_{xy}$ quantized to $h/4e^2$ and $h/3e^2$, respectively. To rule out the possibility of a conventional

quantum Hall effect, we have further investigated the behavior of both states under varying magnetic fields, both display characteristics of the quantum anomalous Hall effect[42–44] (Fig. 2d-e). The $C$ = 4 state shows an unprecedentedly high critical temperature of $T_c$ ~ 8.5 K in graphene (Supplementary Fig. 6a). Unlike the $C$ = 4 state, which can be precisely quantized at zero magnetic field, the $C$ = 3 state requires a magnetic field of $B_\perp$ ~ 100 mT to achieve quantization, likely due to domain fluctuations at zero magnetic field which are commonly observed in graphene devices[34,37]. Figures 2h-i (Figures 2j-k) show the Landau fan diagrams for the $C$ = 4 ($C$ = 3) state. Both states follow the Středa formula and persist at zero magnetic field. In summary, we observe Chern insulators with $C$ = 4 ($C$ = 3) at $\nu$ = 1 under lower (higher) $D$ fields, which is consistent with our Hartree-Fock calculations (Extended Data Fig. 2). Such $C$ = 4 and $C$ = 3 Chern insulators modulated by $D$ field at $\nu$ = 1 can also be observed in device D2 (Extended Data Fig. 3), D3 (Supplementary Fig. 7), D4 (Supplementary Fig. 8) and D5 (Supplementary Fig. 9) with non-interacting band calculations (Supplementary Fig. 5).

**Cascades of high-Chern insulators**

Near $\nu$ = 3, we observe a series of unprecedentedly rich high Chern insulators. Figure 3a shows $\rho_{xx}$ as a function of $D/\varepsilon_0$ and $\nu$ near $\nu$ = 3 with $\rho_{xy}$ shown in Fig. 3b. In the phase diagram, we can clearly identify regions where the $\rho_{xx}$ vanishes, corresponding to constant values of $\rho_{xy}$. Figure 3k quantifies the evolution of $\rho_{xx}$ and $\rho_{xy}$ along the red dashed line in Fig. 3a. A series of Hall plateaus with $|\rho_{xy}|$ = $h/e^2$, $h/2e^2$, $h/3e^2$, $h/4e^2$, $h/5e^2$, $h/6e^2$, $h/7e^2$ can be distinguished, with each corresponding to a minimal or completely vanishing $\rho_{xx}$. We further summarize these high Chern insulator states in the schematic (Fig. 3c). To exclude the possibility that these quantized Hall states originate from Landau levels, we performed measurements under magnetic field. Figures 3d-j display the corresponding magnetic hysteresis loops for these states, all of which exhibit quantum anomalous Hall effect with precise quantization. At high magnetic fields, these Chern insulators show complex fan diagrams (Fig. 3l and Extended Data Fig. 4f-l). These Chern insulators compete in the phase diagram, distinctly differing from the traditional Chern gap and neighboring quantum Hall states.

Moreover, a clear phase boundary between positive and negative Hall resistance can be identified in Fig. 3a–b. This boundary is dependent on the scan direction of $\nu$, when $\nu$ is increased, the $C < 0$ region expands, whereas the $C > 0$ region is favored when $\nu$ is decreased (Extended Data Fig. 4a–b). This is also reflected in the fan diagrams, where different Chern insulators are preferentially stabilized depending on scan direction (Fig. 3l and Extended Data Fig. 4f-l). As a result, a pronounced hysteresis loop develops as a function of $\nu$, manifesting as the direction-dependent phase boundary in the Hall response (Extended Data Fig. 4d–e). This behavior reflects a first-order transition between competing valley-polarized states. The total magnetization receives contributions from the spin magnetization induced by bulk carrier and orbital magnetization induced by edge currents. The two components compete as the carrier density is varied, and can lead to sign reversal of total magnetization, driving valley switching to minimize Zeeman energy[45,46]. Because the valley polarized states are separated by an energy barrier, the system can remain trapped in a metastable state depending on the scan history of $\nu$, giving rise to the observed hysteresis and scan-direction dependence. Similar behavior is observed in device D1 (Extended Data Fig. 5 and Supplementary Fig. 11) and D5 (Supplementary Fig. 10), indicating that it is robust and intrinsic to the system.

We further performed temperature-dependent measurements on these Chern insulator states and determined the energy gaps, ranging from $\Delta \sim 0.25$ meV (for the $C$ = -2 state) to $\Delta \sim 1.63$ meV (for the $C$ = 4 state) (Supplementary Fig. 6). However, the states with $C$ = 5, 6 and 7 exhibit completely anomalous temperature dependence (Fig. 3m-n and Extended Data Fig. 6). As temperature increases, these states deviate from quantization, yet show larger $\rho_{xy}$. This unusual behavior suggests a different origin from traditional quantum anomalous Hall states. Except for the $C$ = 4 state at $\nu$ = 3, other states are continuously distributed in the non-commensurate regions from $\nu$ = 2.5 to $\nu$ = 3.5.

Regarding the unusual temperature dependence of the high-Chern states ($C$ = 5–7), we consider a possible scenario involving an anomalous Hall crystal (AHC)[47–50], in which the quantum anomalous Hall effect coexists with spontaneous translation-symmetry breaking in the form of charge density wave (CDW) order (Extended Data Fig. 7). In such a picture, the system near $\nu \approx 3$ can be viewed as a partially filled electron liquid ($\nu' = \nu-3$) on top of a robust $|C|$ = 4 Chern-insulator background. At low temperatures, this partially filled component may freeze into a CDW state with enlarged periodicity, forming an AHC with additional quantized Berry curvature, yielding the observed higher Chern numbers. Within this framework, the non-monotonic evolution of $\rho_{xy}$ with temperature can be naturally understood. As temperature increases, thermal fluctuations first weaken and melt the CDW (AHC) order, reducing the additional Hall contribution from the non-commensurate component and leaving the $|C|$ = 4 state behind. This leads to an apparent increase in $\rho_{xy}$ beyond the low-temperature quantized values. At higher temperatures, further suppression of spin–valley polarization destabilizes the Chern-insulator background, leading to a dissipative metallic state with decreased $\rho_{xy}$ and increased $\rho_{xx}$. Consistent with this interpretation, we find that increasing the excitation current, which effectively raises the electronic temperature, produces a qualitatively similar two-step evolution in both $\rho_{xx}$ and $\rho_{xy}$ (Supplementary Fig. 12). These observations support a picture where the high-Chern states arise from the coexistence of a conventional Chern insulator and an additional interaction-driven ordered component. We note, however, that a quantitative understanding of this mechanism requires further theoretical investigation beyond the current work.

## Fractional high Chern insulator

More strikingly, we observe the behavior of a FCI at $\nu$ = 2/3 at three independent devices. Figures 4a-c show $\rho_{xx}$ measured from device D2, D1 and D5, as a function of $\nu$ and $D$ field. Figure 4d-f shows the corresponding $\rho_{xy}$. Near $\nu$ = 2/3, clear local minimums in $\rho_{xx}$ and a maximum in $\rho_{xy}$ can be resolved. This can be clearly reflected in Fig. 4g-i, which display the linecut indicated by the horizontal dashed lines in Fig. 4a-c. The local minimum of $\rho_{xx}$ near $\nu$ = 2/3 can be down to 3 kΩ. Importantly, the $\rho_{xy}$ is quantized to $\rho_{xy} = 3h/7e^2$ with 99% accuracy, indicating the formation of an FCI with $C$ = 7/3 (Fig. 4g-i). The FCI state further shows typical anomalous Hall effect and decent quantization of $\rho_{xy} = 3h/7e^2$ with an accuracy of more than 98% when $B_\perp$ > 200mT. To further confirm the FCI nature of this state, we measure its evolution at high magnetic fields. As shown in Fig. 4m-r, both $\rho_{xx}$ and $\rho_{xy}$ of the state strictly follow the Středa formula corresponding to $C$ = 7/3. Notably, the observed FCI state is unexpectedly suppressed at high magnetic field, which is possibly attributed to magnetic field induced distortion of band geometry and has been observed in other moiré systems[16,34] . We also perform temperature-dependent measurements on the state (Extended Data Fig. 8 and Supplementary Fig. 6), which exhibit clear activation behavior with ~ 0.51 meV at $B_\perp = 0.3$ T.

The observation of an FCI state with $C = 7/3$ at $\nu = 2/3$ is highly intriguing. Such an interaction-driven fractionalized phase can occur in a topological flat band, although the Chern number $C$ of the parent band is not yet settled by current experiments and calculations—it could be either $C = 3$ or $C = 4$. This phase is interesting because the many-body Chern number $C_{mb} = 7/3$ is not equal to $\nu C$, which would be 2 (for $C = 3$) or 8/3 (for $C = 4$), placing it outside the single Landau level paradigm.

To illustrate the origin of the $C = 7/3$ fractional Chern insulator state, we propose two possible mechanisms based on different parent bands:

The first proposed possible mechanism for this FCI phase is under the assumption that the parent Chern number is $C = 3$ (Extended Data Fig. 9a). Our key idea is that Hall conductivity $\sigma_{xy} = (7/3)e^2/h$ originates from two groups of electrons: one with filling $\nu^{(1)} = 1/3$ contributes $\sigma_{xy}^{(1)} = 2e^2/h$ through $\sqrt{3} \times \sqrt{3}$ charge ordering, which spontaneously breaks the discrete moiré translation symmetry[51–53]; the remaining electrons, at filling $\nu^{(2)} = 1/3$, fractionalize to contribute $\sigma_{xy}^{(2)} = (1/3)e^2/h$ via the conventional fractional quantum Hall mechanism in a $C = 1$ topological band. More concretely, our mechanism assumes that the $C = 3$ parent band is down-folded into a Chern structure (0,1,2) or (1,0,2) (from high to low energy) when the real-space $\sqrt{3} \times \sqrt{3}$ charge ordering occurs. Filling the lowest folded band occupies a filling of $\nu^{(1)} = 1/3$ and contributes a Hall conductivity of $\sigma_{xy}^{(1)} = 2e^2/h$. Given this charge ordering, integrating out these electrons yields a renormalized topological band with $C = 1$ at filling $\nu^{(2)} = 1/3$, which further fractionalizes into a Laughlin $\nu = 1/3$ fractional Chern insulator.

Another potential mechanism relies on the $C = 4$ parent band, which can be effectively mapped to a four-layer $C = 1$ system via unit cell enlargement[23,25,54,55] (Extended Data Fig. 9b). Specifically, a $4 \times 1$ or $2 \times 2$ charge order can fold the $C = 4$ band into four $C = 1$ bands, such that original filling $\nu = 2/3$ corresponds to $\nu' = 8/3$ in the supercell. In this picture, two $C = 1$ bands are fully occupied by $\nu'^{(1)} = 2$ and one $C = 1$ band is occupied by $\nu'^{(2)} = 2/3$, contributing to $\sigma_{xy}^{(1)} = 2e^2/h$ and $\sigma_{xy}^{(2)} = (2/3)e^2/h$, respectively. To account for the total $\sigma_{xy} = (7/3)e^2/h$, an additional contribution $\sigma_{xy}^{(3)} = -(1/3)e^2/h$ is required. As detailed in the Method, this negative contribution can arise from neutral particle–hole excitations between the two valleys carrying opposite Chern numbers. In particular, configurations involving 1/8 particles and 1/8 holes in the two valleys are described by a Halperin wavefunction. Consequently, the resulting state constitutes a valley-entangled topological phase, in which excitations in the two valleys exhibit nontrivial braiding statistics.

So far, we cannot rule out other mechanisms and more systematic studies combining experiment and many-body calculations are required.

## Conclusion

In summary, we have observed a rich variety of both integer and fractional Chern insulators in a new moiré system composed of Bernal bilayer graphene and rhombohedral tetralayer graphene.

The $C = 7/3$ state at $\nu = 2/3$ is beyond the current theoretical framework for fractional quantum Hall effects as well as the conventional high Chern theory. Additionally, we observe an unprecedented richness of quantum anomalous Hall states at $\nu = 1$ and $\nu = 3$. These high Chern insulators compete and intertwine with each other, greatly expanding our understanding of conventional band topology. Further experimental investigations, incorporating additional tuning parameters such as twist angles and layer numbers, are promising for exploring new fractional charges and potential non-Abelian anyons.

## References

1. Tsui, D. C., Stormer, H. L. & Gossard, A. C. Two-Dimensional Magnetotransport in the Extreme Quantum Limit. *Phys. Rev. Lett.* **48**, 1559–1562 (1982).
2. Willett, R. *et al.* Observation of an even-denominator quantum number in the fractional quantum Hall effect. *Phys. Rev. Lett.* **59**, 1776–1779 (1987).
3. Neupert, T., Santos, L., Chamon, C. & Mudry, C. Fractional Quantum Hall States at Zero Magnetic Field. *Phys. Rev. Lett.* **106**, 236804 (2011).
4. Tang, E., Mei, J.-W. & Wen, X.-G. High-Temperature Fractional Quantum Hall States. *Phys. Rev. Lett.* **106**, 236802 (2011).
5. Sun, K., Gu, Z., Katsura, H. & Das Sarma, S. Nearly Flatbands with Nontrivial Topology. *Phys. Rev. Lett.* **106**, 236803 (2011).
6. Sheng, D. N., Gu, Z.-C., Sun, K. & Sheng, L. Fractional quantum Hall effect in the absence of Landau levels. *Nat. Commun.* **2**, 389 (2011).
7. Regnault, N. & Bernevig, B. A. Fractional Chern Insulator. *Phys. Rev. X* **1**, 021014 (2011).
8. Arovas, D., Schrieffer, J. R. & Wilczek, F. Fractional Statistics and the Quantum Hall Effect. *Phys. Rev. Lett.* **53**, 722–723 (1984).
9. Wen, X. G. Non-Abelian statistics in the fractional quantum Hall states. *Phys. Rev. Lett.* **66**, 802–805 (1991).
10. Bartolomei, H. *et al.* Fractional statistics in anyon collisions. *Science* **368**, 173–177 (2020).
11. Nakamura, J., Liang, S., Gardner, G. C. & Manfra, M. J. Direct observation of anyonic braiding statistics. *Nat. Phys.* **16**, 931–936 (2020).
12. Cai, J. *et al.* Signatures of fractional quantum anomalous Hall states in twisted MoTe2. *Nature* **622**, 63–68 (2023).
13. Zeng, Y. *et al.* Thermodynamic evidence of fractional Chern insulator in moiré MoTe2. *Nature* **622**, 69–73 (2023).
14. Park, H. *et al.* Observation of fractionally quantized anomalous Hall effect. *Nature* **622**, 74–79 (2023).
15. Xu, F. *et al.* Observation of Integer and Fractional Quantum Anomalous Hall Effects in Twisted Bilayer MoTe 2. *Phys. Rev. X* **13**, 031037 (2023).
16. Lu, Z. *et al.* Fractional quantum anomalous Hall effect in multilayer graphene. *Nature* **626**, 759–764 (2024).
17. Lu, Z. *et al.* Extended quantum anomalous Hall states in graphene/hBN moiré superlattices. *Nature* **637**, 1090–1095 (2025).
18. Waters, D. *et al.* Chern Insulators at Integer and Fractional Filling in Moiré Pentalayer Graphene. *Phys. Rev. X* **15**, 011045 (2025).
19. Choi, Y. *et al.* Superconductivity and quantized anomalous Hall effect in rhombohedral graphene. *Nature* **639**, 342–347 (2025).

20. Xie, J. *et al.* Tunable fractional Chern insulators in rhombohedral graphene superlattices. *Nat. Mater.* **24**, 1042–1048 (2025).
21. Liu, Z., Bergholtz, E. J., Fan, H. & Läuchli, A. M. Fractional Chern Insulators in Topological Flat Bands with Higher Chern Number. *Phys. Rev. Lett.* **109**, 186805 (2012).
22. Yang, S., Gu, Z.-C., Sun, K. & Das Sarma, S. Topological flat band models with arbitrary Chern numbers. *Phys. Rev. B* **86**, 241112 (2012).
23. Sterdyniak, A., Repellin, C., Bernevig, B. A. & Regnault, N. Series of Abelian and non-Abelian states in C > 1 fractional Chern insulators. *Phys. Rev. B* **87**, 205137 (2013).
24. Möller, G. & Cooper, N. R. Fractional Chern Insulators in Harper-Hofstadter Bands with Higher Chern Number. *Phys. Rev. Lett.* **115**, 126401 (2015).
25. Wang, J. & Liu, Z. Hierarchy of Ideal Flatbands in Chiral Twisted Multilayer Graphene Models. *Phys. Rev. Lett.* **128**, 176403 (2022).
26. Zhang, Y.-H., Mao, D., Cao, Y., Jarillo-Herrero, P. & Senthil, T. Nearly flat Chern bands in moiré superlattices. *Phys. Rev. B* **99**, 075127 (2019).
27. Ledwith, P. J., Vishwanath, A. & Khalaf, E. Family of Ideal Chern Flatbands with Arbitrary Chern Number in Chiral Twisted Graphene Multilayers. *Phys. Rev. Lett.* **128**, 176404 (2022).
28. Repellin, C. & Senthil, T. Chern bands of twisted bilayer graphene: Fractional Chern insulators and spin phase transition. *Phys. Rev. Res.* **2**, 023238 (2020).
29. Ledwith, P. J., Tarnopolsky, G., Khalaf, E. & Vishwanath, A. Fractional Chern insulator states in twisted bilayer graphene: An analytical approach. *Phys. Rev. Res.* **2**, 023237 (2020).
30. Aronson, S. H. *et al.* Displacement Field-Controlled Fractional Chern Insulators and Charge Density Waves in a Graphene/hBN Moiré Superlattice. *Phys. Rev. X* **15**, 031026 (2025).
31. Xie, J. *et al.* Unconventional Orbital Magnetism in Graphene-based Fractional Chern Insulators. Preprint at https://doi.org/10.48550/arXiv.2506.01485 (2025).
32. Huo, Z. *et al.* Does Moire Matter? Critical Moire Dependence with Quantum Fluctuations in Graphene Based Integer and Fractional Chern Insulators. Preprint at https://doi.org/10.48550/arXiv.2510.15309 (2025).
33. Dong, J. *et al.* Observation of Integer and Fractional Chern insulators in high Chern number flatbands. Preprint at https://doi.org/10.48550/arXiv.2507.09908 (2025).
34. Chen, G. *et al.* Tunable correlated Chern insulator and ferromagnetism in a moiré superlattice. *Nature* **579**, 56–61 (2020).
35. Han, T. *et al.* Correlated insulator and Chern insulators in pentalayer rhombohedral-stacked graphene. *Nat. Nanotechnol.* **19**, 181–187 (2024).
36. Han, T. *et al.* Large quantum anomalous Hall effect in spin-orbit proximitized rhombohedral graphene. *Science* **384**, 647–651 (2024).
37. Sha, Y. *et al.* Observation of a Chern insulator in crystalline ABCA-tetralayer graphene with spin-orbit coupling. *Science* **384**, 414–419 (2024).
38. Wang, W. *et al.* Programmable Quantum Anomalous Hall Insulator in Twisted Crystalline Flatbands. *Phys. Rev. X* **16**, 011015 (2026).
39. Liu, N. *et al.* Diverse high-Chern-number quantum anomalous Hall insulators in twisted rhombohedral graphene. Preprint at https://doi.org/10.48550/arXiv.2507.11347 (2025).
40. Cao, Y. *et al.* Correlated insulator behaviour at half-filling in magic-angle graphene superlattices. *Nature* **556**, 80–84 (2018).
41. Chen, S. *et al.* Electrically tunable correlated and topological states in twisted monolayer–bilayer graphene. *Nat. Phys.* **17**, 374–380 (2021).

42. Sharpe, A. L. *et al.* Emergent ferromagnetism near three-quarters filling in twisted bilayer graphene. *Science* **365**, 605–608 (2019).
43. Serlin, M. *et al.* Intrinsic quantized anomalous Hall effect in a moiré heterostructure. *Science* **367**, 900–903 (2020).
44. Bultinck, N., Chatterjee, S. & Zaletel, M. P. Mechanism for Anomalous Hall Ferromagnetism in Twisted Bilayer Graphene. *Phys. Rev. Lett.* **124**, 166601 (2020).
45. Zhu, J., Su, J.-J. & MacDonald, A. H. Voltage-Controlled Magnetic Reversal in Orbital Chern Insulators. *Phys. Rev. Lett.* **125**, 227702 (2020).
46. Polshyn, H. *et al.* Electrical switching of magnetic order in an orbital Chern insulator. *Nature* **588**, 66–70 (2020).
47. Dong, Z., Patri, A. S. & Senthil, T. Stability of anomalous Hall crystals in multilayer rhombohedral graphene. *Phys. Rev. B* **110**, 205130 (2024).
48. Zhou, B., Yang, H. & Zhang, Y.-H. Fractional Quantum Anomalous Hall Effect in Rhombohedral Multilayer Graphene in the Moiréless Limit. *Phys. Rev. Lett.* **133**, 206504 (2024).
49. Dong, J. *et al.* Anomalous Hall Crystals in Rhombohedral Multilayer Graphene. I. Interaction-Driven Chern Bands and Fractional Quantum Hall States at Zero Magnetic Field. *Phys. Rev. Lett.* **133**, 206503 (2024).
50. Soejima, T. *et al.* Anomalous Hall crystals in rhombohedral multilayer graphene. II. General mechanism and a minimal model. *Phys. Rev. B* **110**, 205124 (2024).
51. Xie, Y. *et al.* Fractional Chern insulators in magic-angle twisted bilayer graphene. *Nature* **600**, 439–443 (2021).
52. Dong, J., Ledwith, P. J., Khalaf, E., Lee, J. Y. & Vishwanath, A. Many-body ground states from decomposition of ideal higher Chern bands: Applications to chirally twisted graphene multilayers. *Phys. Rev. Res.* **5**, 023166 (2023).
53. Polshyn, H. *et al.* Topological charge density waves at half-integer filling of a moiré superlattice. *Nat. Phys.* **18**, 42–47 (2022).
54. Barkeshli, M. & Qi, X.-L. Topological Nematic States and Non-Abelian Lattice Dislocations. *Phys. Rev. X* **2**, 031013 (2012).
55. Wang, J., Klevtsov, S. & Liu, Z. Origin of model fractional Chern insulators in all topological ideal flatbands: Explicit color-entangled wave function and exact density algebra. *Phys. Rev. Res.* **5**, 023167 (2023).

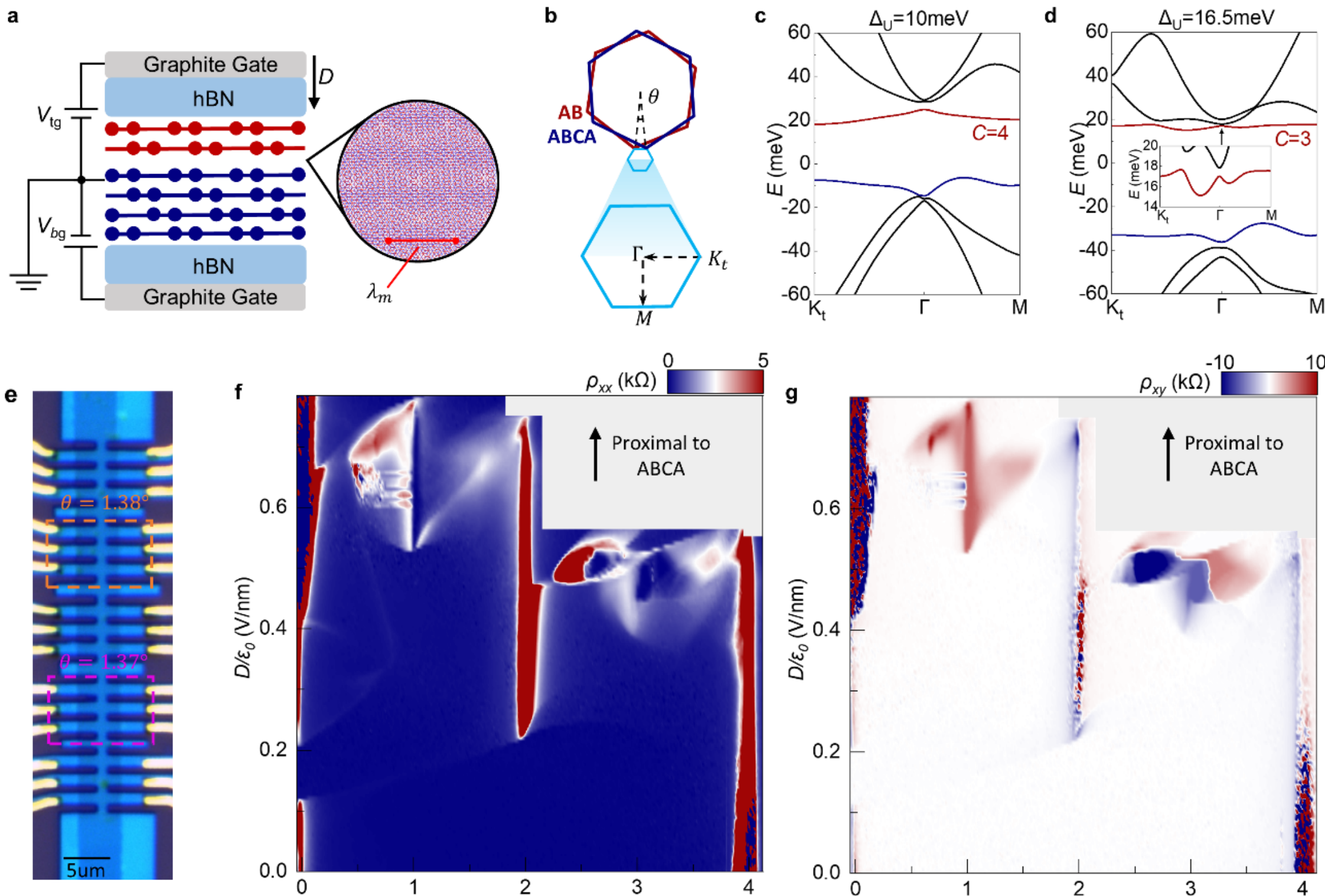

**Fig. 1 | Topological flat band in the twisted bilayer rhombohedral tetralayer graphene. a,** Schematic of the TBRTG device, in which rotationally misaligned Bernal bilayer graphene and rhombohedral stacked tetra-layer graphene with twist angle $\theta$ are sandwiched between two graphite gates with hBN insulating spacer layers, forming a moiré superlattice with lattice period $\lambda_m$. **b,** Schematic of the Brillouin zone of TBRTG, the dashed arrows denote the projection path. **c-d,** Non-interacting band structure of TBRTG at $\theta = 1.38°$ obtained from continuum model with $\Delta_U = 10$ meV **(c)** and $\Delta_U = 16.5$ meV **(d)**, where $\Delta_U$ is the potential difference between adjacent graphene layers. The dark red lines represent the central flat conduction bands with their valley Chern number denoted by $C$. **e,** Optical image of the fabricated sample, which consists of device D1 (orange dashed rectangle) with twist angle $\theta = 1.38°$ and device D2 (purple dashed rectangle) with twist angle $\theta = 1.37°$. **f-g,** Symmetrized longitudinal resistivity $\rho_{xx}$ **(f)** and anti-symmetrized Hall resistivity $\rho_{xy}$ **(g)** as functions of $\nu$ and $D/\varepsilon_0$ measured at perpendicular magnetic field $B_\perp = \pm 0.1$ T and temperature $T = 10$ mK for device D1.

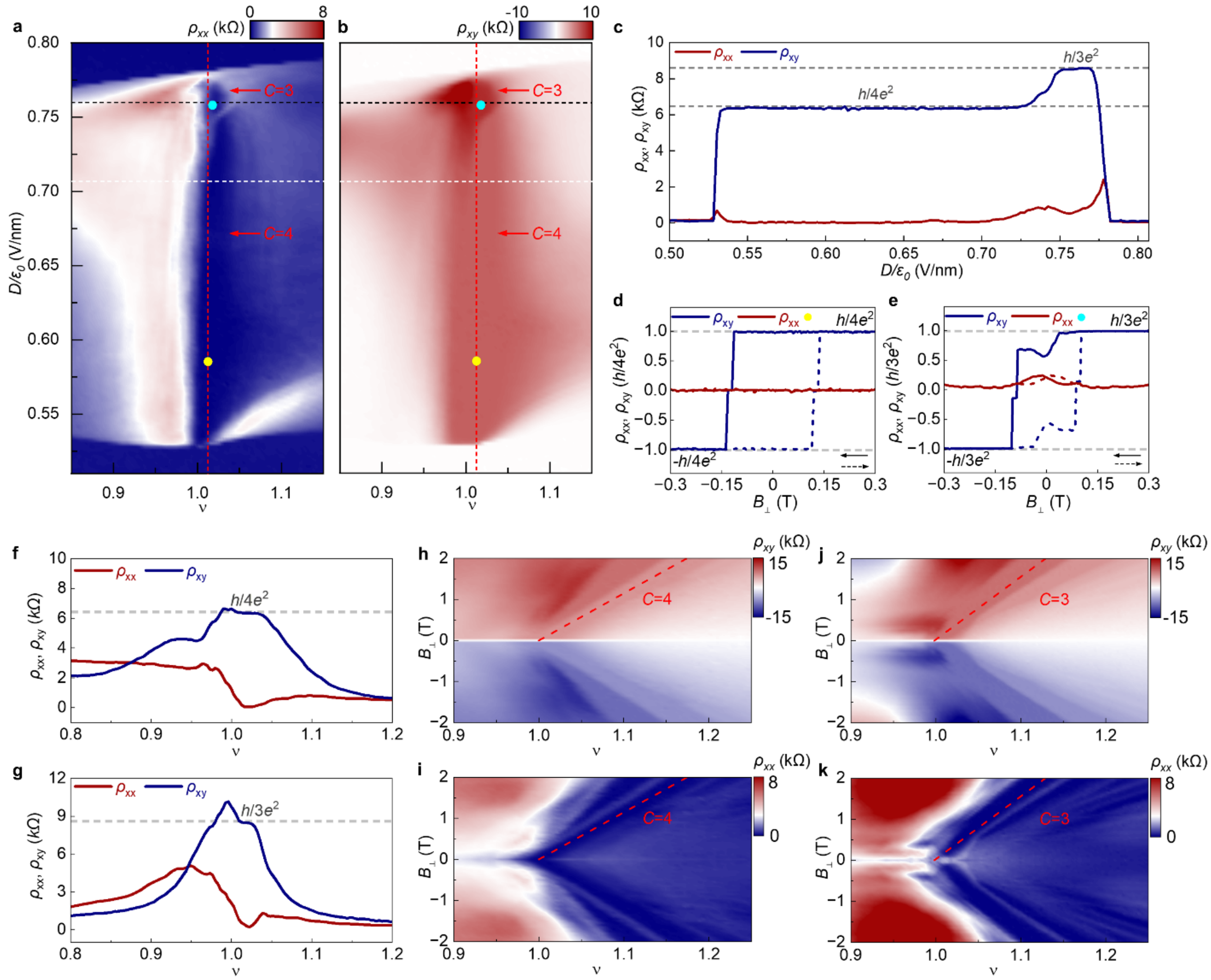

**Fig. 2 | Displacement field modulated Chern insulators at $\nu = 1$. a-b,** Symmetrized $\rho_{xx}$ **(a)** and anti-symmetrized $\rho_{xy}$ **(b)** as functions of $\nu$ and $D/\varepsilon_0$ measured at perpendicular magnetic field $B_\perp = \pm 0.2$ T and temperature $T = 10$ mK around $\nu = 1$ for device D1. The red dashed line is the linecut for **c**. The white dashed line is the linecut for **f**, **h** and **i**. The black dashed line is the linecut for **g**, **j** and **k**. The yellow and cyan circles represent the positions for **d** and **e** respectively. **c,** Symmetrized $\rho_{xx}$ and anti-symmetrized $\rho_{xy}$ as functions of $D/\varepsilon_0$ at $\nu = 1.01$ and $B_\perp = \pm 0.2$ T. **d-e,** Hysteresis loop of Symmetrized $\rho_{xx}$ and anti-symmetrized $\rho_{xy}$ as functions of $B_\perp$ at $\nu = 1.01$, $D/\varepsilon_0 = 0.586$ V/nm **(d)** and $\nu = 1.01$, $D/\varepsilon_0 = 0.757$ V/nm **(e)**. **f-g,** Symmetrized $\rho_{xx}$ and anti-symmetrized $\rho_{xy}$ as a function of $\nu$ at $B_\perp = \pm 0.2$ T for $D/\varepsilon_0 = 0.707$ V/nm **(f)** and 0.757 V/nm **(g)**. **h-k,** Symmetrized $\rho_{xx}$ and anti-symmetrized $\rho_{xy}$ as functions of $\nu$ and $B_\perp$ at $D/\varepsilon_0 = 0.707$ V/nm **(h-i)** and 0.757 V/nm **(j-k)**. The red dashed lines indicate the $\rho_{xx}$ minimal and $\rho_{xy}$ plateaus, which strictly follows the Středa formula $C = h/e(\partial n/\partial B)$.

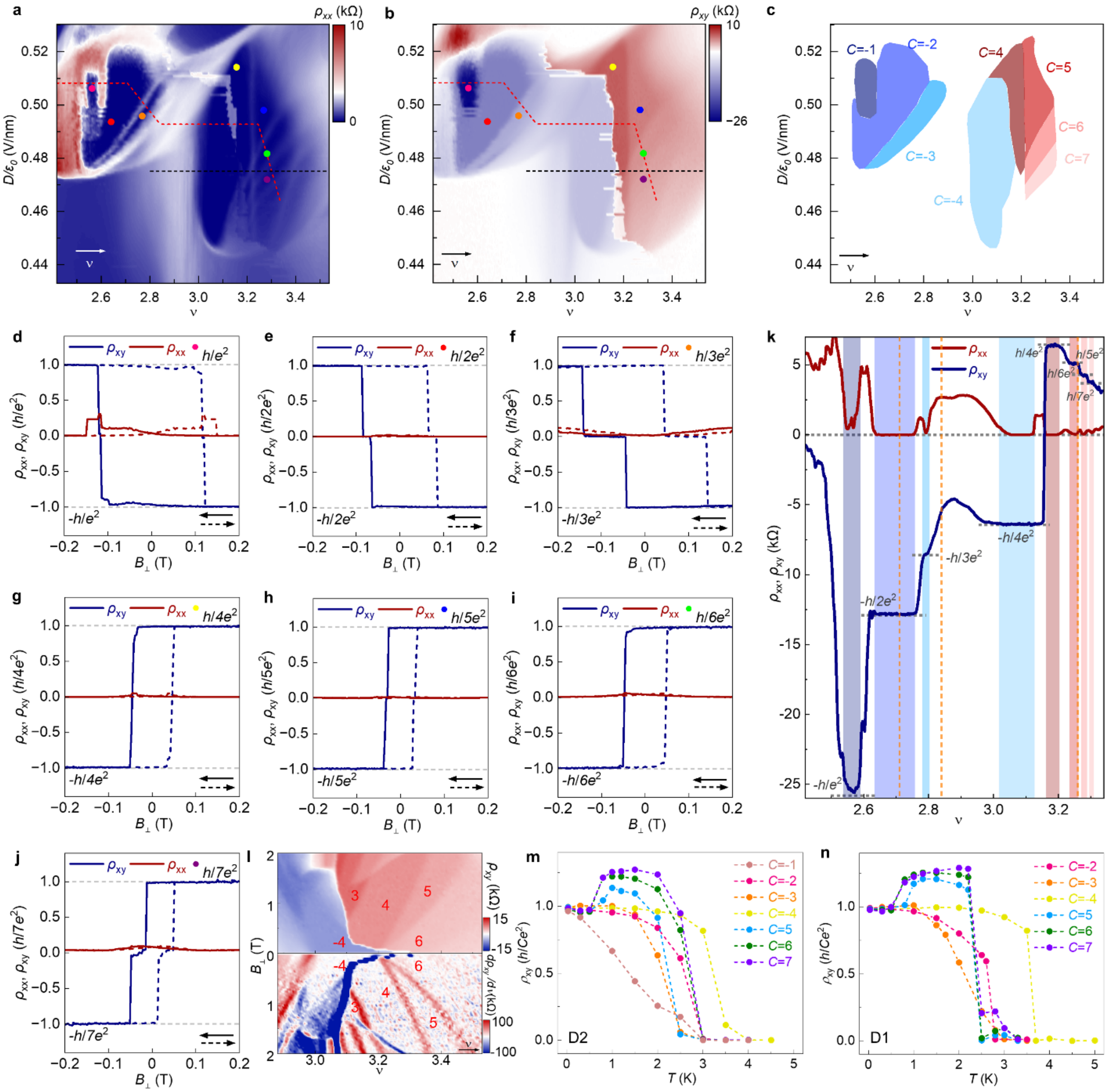

**Fig. 3 | Cascades of high Chern insulators around $\nu = 3$. a-b,** Symmetrized $\rho_{xx}$ **(a)** and anti-symmetrized $\rho_{xy}$ **(b)** as functions of $\nu$ and $D/\varepsilon_0$ measured at perpendicular magnetic field $B_\perp = \pm 0.1$ T and temperature $T = 10$ mK with positive $\nu$ sweeping direction for device D2. The red and black dashed lines represent the linecuts for **k** and **l** respectively. The colored circles represent the positions for **d-j**. **c,** Schematic of the phase diagram of **a-b**, where the Chern insulators with different Chern numbers are highlighted. **d-j,** Hysteresis loops of Symmetrized $\rho_{xx}$ and anti-symmetrized $\rho_{xy}$ as functions of $B_\perp$ at $\nu = 2.56$, $D/\varepsilon_0 = 0.506$ V/nm **(d),** $\nu = 2.64$, $D/\varepsilon_0 = 0.494$ V/nm **(e)**, $\nu = 2.77$, $D/\varepsilon_0 = 0.496$ V/nm **(f)**, $\nu = 3.16$, $D/\varepsilon_0 = 0.514$ V/nm **(g)**, $\nu = 3.27$, $D/\varepsilon_0 = 0.498$ V/nm **(h)**, $\nu = 3.28$, $D/\varepsilon_0 = 0.482$ V/nm **(i)**, and $\nu = 3.28$, $D/\varepsilon_0 = 0.472$ V/nm **(j)**. **k,** Symmetrized $\rho_{xx}$ and anti-symmetrized $\rho_{xy}$ as functions of $\nu$ following the red dashed trace in **a**. Orange dashed lines indicate the positions of the turning points of the red dashed line. **l,** Anti-symmetrized $\rho_{xy}$ and derivative of anti-symmetrized $\rho_{xy}$ on $\nu$ as functions of $\nu$ and $B_\perp$ at $D/\varepsilon_0 = 0.475$ V/nm with positive

$\nu$ sweeping direction. **m-n,** Normalized and anti-symmetrized $\rho_{xy}$ at $B_\perp = 0$ T as a function of temperature for device D2 **(m)** and device D1 **(n)** extracted from Extended Data Fig. 6 and Supplementary Fig. 6. The colors represent Chern insulators around $\nu = 3$ with different Chern numbers.

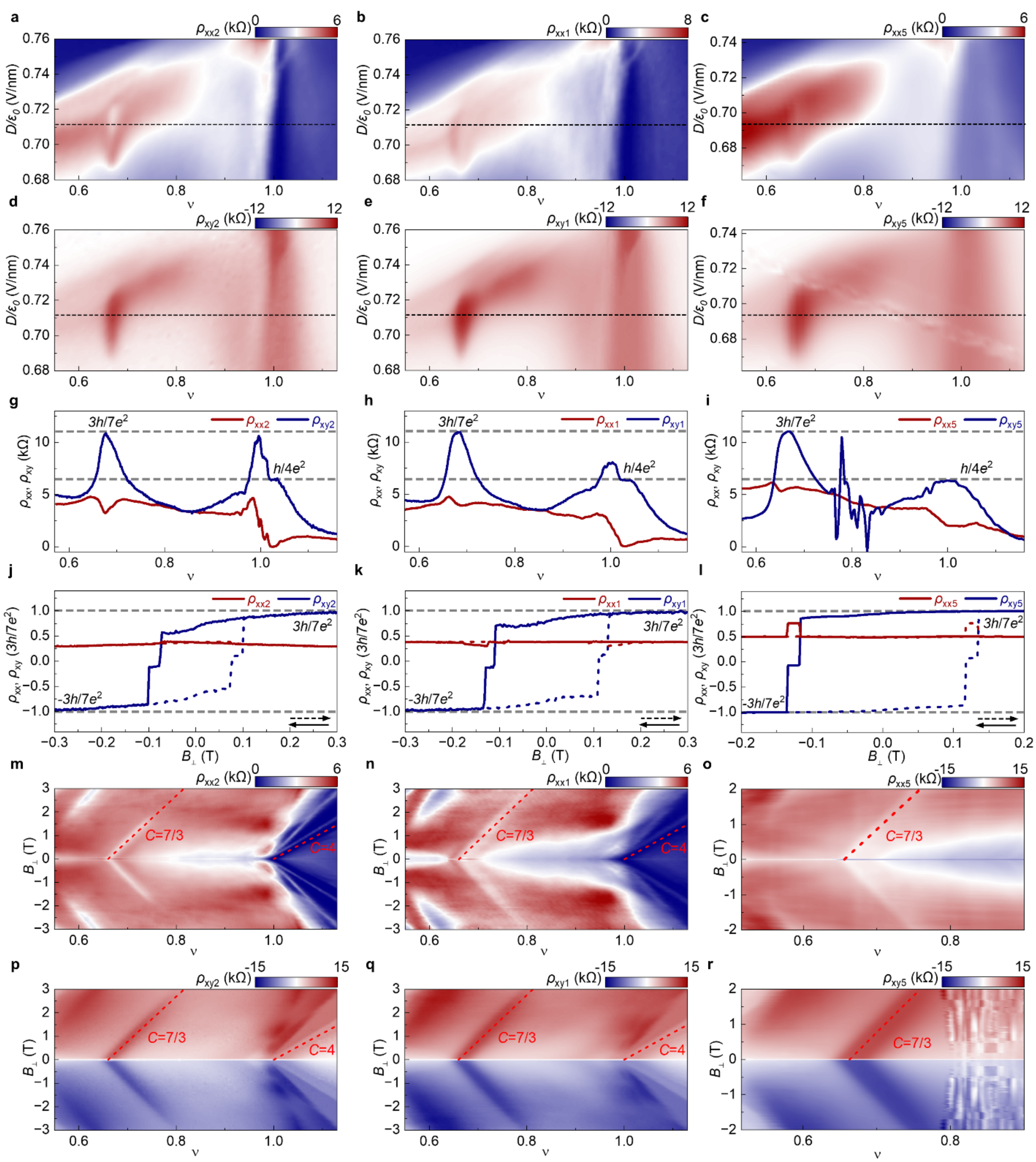

**Fig. 4 | Fractional Chern insulator at $\nu$ = 2/3. a-c,** Symmetrized $\rho_{xx}$ as functions of $\nu$ and $D/\varepsilon_0$ measured at perpendicular magnetic field $B_\perp = \pm 0.1$ T and temperature $T = 10$ mK for device D2 **(a)**, D1 **(b)** and D5 **(c)**. **d-f,** Anti-symmetrized $\rho_{xy}$ as functions of $\nu$ and $D/\varepsilon_0$ measured at perpendicular magnetic field $B_\perp = \pm 0.1$ T and temperature $T = 10$ mK for device D2 **(d)**, D1 **(e)**

and D5 **(f)**. The black dashed lines are the linecuts for **g-i** and **m-r**. **g-i,** Symmetrized $\rho_{xx}$ and anti-symmetrized $\rho_{xy}$ as a function of $\nu$ at $B_\perp = \pm 0.3$ T with $D/\varepsilon_0 = 0.711$ V/nm for device D2 **(g)** and D1 **(h)**, and $B_\perp = \pm 0.1$ T with $D/\varepsilon_0 = 0.694$ V/nm for device D5 **(g)**. **j-l,** Hysteresis loop of Symmetrized $\rho_{xx}$ and anti-symmetrized $\rho_{xy}$ as functions of $B_\perp$ at $\nu = 2/3$, $D/\varepsilon_0 = 0.713$ V/nm for device D2 **(j)** and D1 **(k)**, and $D/\varepsilon_0 = 0.694$ V/nm for device D5**(l)**. **m-o,** Symmetrized $\rho_{xx}$ as functions of $\nu$ and $B_\perp$ at $D/\varepsilon_0 = 0.711$ V/nm for device D2 **(m)** and D1 **(n)**, $D/\varepsilon_0 = 0.694$ V/nm for device D5 **(o)**. **p-r,** anti-symmetrized $\rho_{xy}$ as functions of $\nu$ and $B_\perp$ at $D/\varepsilon_0 = 0.711$ V/nm for device D2 **(p)** and device D1 **(q)**, $D/\varepsilon_0 = 0.694$ V/nm for device D5 **(r)**. The red dashed lines indicate the $\rho_{xx}$ minimal and $\rho_{xy}$ plateaus, which strictly follows the Středa formula $C = h/e(\partial n/\partial B)$.

# Methods

## Device Fabrication

The devices consist of bilayer graphene and rhombohedral tetralayer graphene encapsulated between two hBN flakes, with graphite used as both the top and bottom gate. Graphene and hBN (10-20 nm thick) flakes were exfoliated onto $O_2$ plasma cleaned $SiO_2$/Si substrate. Bilayer and tetralayer graphene were identified by their optical contrast with substrate under a microscope, and the rhombohedral region of tetralayer graphene was identified via infrared imaging technique[56] and further confirmed by Raman spectroscopy. After that, the rhombohedral tetralayer graphene was segmented into several isolated pieces using a femtosecond laser (HR-Femto-Sci, with a central wavelength ~517 nm, pulse width ~ 150 fs, repetition frequency ~ 80 MHz and a maximum power ~150 mW). A PC (poly bisphenol A carbonate)/PDMS (polydimethylsiloxane) stamp was used to first pick up an hBN flake, followed by the sequential pickup of bilayer graphene and rhombohedral tetralayer graphene at controlled twist angles. The assembled stack was then released onto a pre-fabricated hBN/graphite gate substrate that had been cleaned by the contact mode of atom force microscope. Finally, the top graphite gate was subsequently aligned and transferred onto the stack after cleaning the top hBN surface with atomic force microscopy. After the transfer, Raman spectroscopy was used to verify that the tetralayer graphene retained rhombohedral stacking (Extended Data Fig. 1c). The samples were then patterned into Hall bar geometry by electron beam lithography and ion etched in $CHF_3/O_2$ atmosphere. Electrical connections were established by edge contact of Cr/Au (5 nm/60 nm) electrodes.

## Electrical transport measurement

Low-temperature transport measurements were performed in a dilution refrigerator with a base temperature of approximately 30 mK for all devices. Standard low-frequency lock-in techniques were employed, with gate voltages supplied by Keithley 2400 source meters and an excitation current of 1 nA at 17.777 Hz generated using SR860 lock-in amplifier. The detailed contact configurations of the devices are shown in Extended Data Fig. 1d,f.

## Twist angle determination

By independently tuning the top- and bottom-gate voltages $V_{tg}$ and $V_{bg}$, we control both the carrier density n and electrical displacement field D, which are related through $n = (C_{tg}V_{tg} + C_{bg}V_{bg})/e$ and $D/\varepsilon_0 = (C_{tg}V_{tg} - C_{bg}V_{bg})/2\varepsilon_0$. Here, $C_{tg}$ and $C_{bg}$ denote the capacitances per unit area of the respective gates, and are obtained from the slopes of the Landau-fan diagrams (Supplementary Fig. 1a-b). The twist angles are inferred from the carrier density at filling factor $\nu = 4$, and are consistent with those independently obtained from Brown–Zakk oscillations (Supplementary Fig. 1c-f).

**Chern number determination**

The Chern numbers are determined from the evolution of the longitudinal resistance $\rho_{xx}$ with magnetic field. For each state, the positions of the $\rho_{xx}$ minima or $\rho_{xy}$ maximum are identified as a function of carrier density $n$ and perpendicular magnetic field $B_\perp$. The slopes of these trajectories are extracted by linear fitting and are directly related to the Chern number via the Středa formula, $C = (h/e)\, \partial n/\partial B$. In practice, we perform linear fits to the $\rho_{xx}$ minima and determine $\partial n/\partial B$, from which the corresponding Chern numbers are obtained. Representative fittings are shown in Supplementary Fig. 2.

**Symmetrization and anti-symmetrization**

Since the longitudinal resistivity $\rho_{xx}$ and Hall resistivity $\rho_{xy}$ are expected to be symmetric and antisymmetric with respect to the magnetic field $B$, respectively, we symmetrize and anti-symmetrize the raw data as follows:

$$\rho_{xx}^{Sym}(B,\nu) = \frac{\rho_{xx}^{Original}(B,\nu) + \rho_{xx}^{Original}(-B,\nu)}{2},$$

$$\rho_{xy}^{AntiSym}(B,\nu) = \frac{\rho_{xy}^{Original}(B,\nu) - \rho_{xy}^{Original}(-B,\nu)}{2}.$$

Similarly, for the magnetic hysteresis measurement data, we applied analogous symmetrization and anti-symmetrization procedures:

$$\rho_{xx,Forward}^{Sym}(B) = \frac{\rho_{xx,Forward}^{Original}(B) + \rho_{xx,Backward}^{Original}(-B)}{2},$$

$$\rho_{xx,Backward}^{Sym}(B) = \rho_{xx,Forward}^{Sym}(-B),$$

$$\rho_{xy,Forward}^{AntiSym}(B) = \frac{\rho_{xy,Forward}^{Original}(B) - \rho_{xy,Backward}^{Original}(-B)}{2},$$

$$\rho_{xy,Backward}^{AntiSym}(B) = -\rho_{xy,Forward}^{AntiSym}(-B).$$

**Continuum Model and Interaction**

The free Hamiltonian $\widehat{H}_0$ of TBRTG consists of the kinetic energy of the top and bottom layers $\widehat{H}_t$, $\widehat{H}_b$, a moiré coupling $\widehat{H}_m$, and the displacement field $\widehat{H}_D$. In the $\eta = +$ valley, for both spin $s = \uparrow,\downarrow$,

$$\widehat{H}_0 = \widehat{H}_t + \widehat{H}_b + \widehat{H}_m + \widehat{H}_D$$

$$\widehat{H}_t = \sum_{k,s} \sum_{Q\in Q_t} \Big( \hbar v_F \sum_{l=1}^{2} \psi_{k,Q,l,A}^{\dagger}\, (k-Q)_- \psi_{k,Q,l,B} + \gamma_1 \psi_{k,Q,1,B}^{\dagger} \psi_{k,Q,2,A}$$

$$+\hbar v_3 \psi_{k,Q,1,A}^{\dagger} (k-Q)_+ \psi_{k,Q,2,B} + \hbar v_4 \sum_{\alpha} \psi_{k,Q,1,\alpha}^{\dagger}\, (k-Q)_- \psi_{k,\ \ ,2,\alpha} \Big) + H.c.$$

$$\widehat{H}_b = \sum_{k} \sum_{Q\in Q_b} \Big( \hbar v_F \sum_{l=1}^{4} \psi_{k,Q,l,A}^{\dagger}\, (k-Q)_- \psi_{k,Q,l,B} + \gamma_1 \sum_{l=1}^{3} \psi_{k,Q,l,A}^{\dagger}\, \psi_{k,Q,l+1,B} + \gamma_2 \sum_{l=1}^{2} \psi_{k,Q,l,B}^{\dagger}\ \psi_{k,Q,l+2,A}$$

$$+\hbar v_3 \sum_{l=1}^{3} \psi_{k,Q,l,B}^{\dagger}\, (k-Q)_- \psi_{k,Q,l+1,A} + \hbar v_4 \sum_{l=1}^{3} \sum_{\alpha} \psi_{k,Q,l,\alpha}^{\dagger}\, (k-Q)_+ \psi_{k,Q,l+1,\alpha} \Big) + H.c.,$$

$$\hat{H}_m = w_1 \sum_{k} \sum_{Q \in Q_t} \left( \sum_{d=1}^{3} \sum_{\alpha,\alpha'} \psi^{\dagger}_{k,Q,1,\alpha'} \, [T_d]_{\alpha',\alpha} \psi_{k,Q-q_d,1,\alpha} + H.c. \right),$$

$$T_d = \begin{pmatrix} u_0 & e^{-i\frac{2\pi}{3}(d-1)} \\ e^{i\frac{2\pi}{3}(d-1)} & u_0 \end{pmatrix}.$$

$\psi_{k,Q,l,\alpha,\eta,s}$ with $Q \in Q_t(Q_b)$ denotes an electron of layer $l \in [1,2]$ ($l \in [1,4]$), sublattice $\alpha = A, B$, valley $\eta = \pm$, spin $s = \uparrow,\downarrow$, in the top (bottom) stack. Valley $\eta$ and spin $s$ together form a 4-dimensional flavor space. In the above, we have written $\psi_{k,Q,l,\alpha,+,s} = \psi_{k,Q,l,\alpha}$ for brevity. The moiré Brillouin zone is set by $k_\theta = \theta \cdot \frac{4\pi}{3a_0}$, where $\theta = 1.38°$ is the twist angle, and $a_0 = 0.246$ nm is the graphene lattice constant. $a_M = \frac{a_0}{\theta}$ is the moiré lattice constant. $q_d = k_\theta \left( cos\frac{2\pi(d-1)}{3}, sin\frac{2\pi(d-1)}{3} \right)$ ($d = 1,2,3$). The moiré reciprocal lattice is spanned by $b_1 = q_2 - q_1$ and $b_2 = q_3 - q_1$, and $Q_{t/b} = \{m_1 b_1 + m_2 b_2 \mp q_1 | m_{1,2} \in Z\}$, respectively. We admit the parameters measured in Ref [57], i.e., $\hbar v_F = 574$ meV·nm, $\gamma_1 = 364$ meV, $\gamma_2 = -6$ meV, $\hbar v_3 = -0.037\, \hbar v_F$, $\hbar v_4 = -0.047\, \hbar v_F$. $w_1 = 110$ meV, $u_0 = 0.8$. We also define $\psi_{k+G,Q+G,l,\alpha} = \psi_{k,Q,l,\alpha}$, for any moiré reciprocal lattice $G$.

The displacement field is modeled by

$$\hat{H}_D = \Delta_U \sum_{k} \sum_{\alpha} \left( \sum_{l=1}^{2} \left(l - \frac{1}{2}\right) \sum_{Q \in Q_t} \psi^{\dagger}_{k,Q,l,\alpha} \, \psi_{k,Q,l,\alpha} - \sum_{l=1}^{4} \left(l - \frac{1}{2}\right) \sum_{Q \in Q_b} \psi^{\dagger}_{k,Q,l,\alpha} \, \psi_{k,Q,l,\alpha} \right)$$

where $\Delta_U$ denotes the energy difference between adjacent layers.

The free Hamiltonian in the other valley $\eta = -$ can be obtained by the (spinless) time-reversal symmetry $T$, which acts as $T\psi_{k,Q,l,\alpha,\eta,s}T^{-1} = \psi_{-k,-Q,l,\alpha,\overline{\eta},s}$.

Numerical results show that $\hat{H}_0$ at $\Delta_U = 0$ meV has a direct gap ~1.8 meV and an isolated $C = 3$ band above the charge neutrality point (CNP). The CNP gap closes around $\Delta_U = 3.25$ meV and, as $\Delta_U$ increases, reopens with an isolated $C = 4$ band above CNP. This is consistent with the resistivity measurement Fig. 1(f-g) where a gap closure happens at $\nu = 0$, $D/\varepsilon_0 \approx 0.18$ V/nm. As $\Delta_U$ further increases across 15.35 meV, the gap between the first and second bands above CNP closes and reopens with changing the first band's Chern number from 4 to 3. In the following section, our Hartree-Fock calculation will demonstrate a full spin-valley polarization at $\nu = 1$, and hence, one can expect the $C = 4 \rightarrow 3$ transition to be observed at $\nu = 1$. Also, suppose $\Delta_U$ in continuum model is proportional to experimental displacement field $D$, the first gap closure with $\Delta_U = 3.25$ meV, $D/\varepsilon_0 \approx 0.18$ V/nm suggests that the $C = 4 \rightarrow 3$ transition at $\Delta_U = 15.35$ meV would happen around $D/\varepsilon_0 \approx 0.78$ V/nm and $\nu = 1$. See Fig. 2c, there is indeed a transition between $C = 4$ and $C = 3$ Chern insulators at $D/\varepsilon_0 \approx 0.74$ V/nm and $\nu \approx 1$, which is remarkably close to the predicted position.

As shown in Supplementary Fig. 4a, the first conduction band with $|C| = 4$ approximately satisfies the trace condition, which is known to stabilize FCI states and to enable a mapping of the parent Chern band onto $|C|$ intertwined lowest Landau levels[73]. Motivated by this observation, we invoke the Landau-level analogy to propose the two possible mechanisms for the observed FCI states discussed in the main text.

We also note that, as a consequence of lacking mirror symmetry, the first conduction band with $D < 0$, though possess nonvanishing Chern number per spin and valley, is more dispersive (Supplementary Fig. 4b). Therefore, the Fock exchange interaction is not sufficiently dominant to induce spin- and valley-polarization and to open a spectral gap with nonvanishing Chern number. This is consistent with the observed metallic region on the negative-$D$ side over $0 < \nu < 4$ (Supplementary Fig. 3a).

We take the Coulomb interaction as the 2D double-gated form,

$$\hat{H}_{int} = \frac{1}{2N_M}\sum_{q,G}\frac{V(q+G)}{\Omega_M}\delta\hat{\rho}_{q+G}\delta\hat{\rho}_{-q-G},$$

$$\delta\hat{\rho}_{q+G} = \sum_{l,\alpha,\eta,s}\left(\psi^{\dagger}_{k+q,Q-G,l,\alpha,\eta,s}\psi_{k,Q,l,\alpha,\eta,s} - \langle\hat{\rho}_{q+G}\rangle_{ref}\,\delta_{q,0}\right)$$

where $\Omega_M = \frac{\sqrt{3}}{2}a_M^2$ is the area of a moiré unit cell, and $N_M$ denotes the number of moiré unit cells. $V(q) = \pi\xi^2 U_\xi \frac{\tanh\frac{\xi|q|}{2}}{\frac{\xi|q|}{2}}$, where $\xi$ is the distance between two metallic gates, and $U_\xi = \frac{e^2}{4\pi\epsilon\xi}$. $\langle\hat{\rho}_{q+G}\rangle_{ref}$ denotes the expectation value of $\hat{\rho}_{q+G}$ over the reference state where all the non-interacting bands below the charge neutrality point is occupied. Hence, $\delta\hat{\rho}$ is the density deviation from the reference state background. For a dielectric constant $\varepsilon \approx 6$ and screening length $\xi = 10$ nm, we have $U_\xi \approx 24$ meV.

**Hartree-Fock calculations**

We perform self-consistent Hartree-Fock (HF) calculations based on the continuum model introduced above. For each valley $\eta = \pm$ and spin $s = \uparrow,\downarrow$, we retain a subspace consisting of 2 non-interacting bands above and 1 band below the CNP (including more bands will not qualitatively change the results). The screened Coulomb interaction $\hat{H}_{int}$ is projected into this subspace. Since in most of the region we concern (8 meV $< \Delta_U <$ 16meV, 0.40 V/nm $< D/\varepsilon_0 <$ 0.80 V/nm) the first band above CNP (per flavor) is well separated from other bands, we refer it as the active band. The bands further beyond CNP are called remote conduction bands, and bands below CNP are called remote valence bands. Hence, our HF calculations retain 1 remote valence band, 1 active band, and 1 remote conduction band for each flavor ($3 \times 4 = 12$ bands in total).

Denote the noninteracting Bloch states as $c^{\dagger}_{n,\eta,s}(k)|0\rangle$, where $|0\rangle$ is the vacuum state and $n \in [1,3]$ is the band index. The HF order parameter is the band- and flavor-resolved correlations $\{R^{\eta s,\eta' s'}_{nm}(k) \equiv \langle c^{\dagger}_{n,\eta,s}(k)\, c_{m,\eta',s'}(k)\rangle\}$, where $\langle\ldots\rangle$ denotes expectation value of the occupied HF states under a given filling $\nu$. For the reference state, we have $(R_{ref})^{\eta s,\eta' s'}_{nm}(k) = \delta_{n,1}\delta_{n,m}\delta_{\eta,\eta'}\delta_{s,s'}$. We discretize the moiré Brillouin zone by a $N_k \times N_k$ mesh with $N_k = 15$ (larger $N_k$ will not qualitatively change the results). A small Fermi-Dirac smearing temperature $k_B T_{HF} = k_B \times 30$ mK = 0.0026 meV is introduced to stabilize convergence, chosen to be smaller than the typical single-particle level spacing between adjacent **k**-points. Note that the smearing $k_BT_{HF}$ is also consistent with the experimental temperature.

At each HF iteration, we compute the HF self-energy $\sum[R]$, and the effective Hamiltonian $H_{eff}(k) = H_0(k) + \sum[R](k) - \sum_{ref}(k)$, where $\sum_{ref} = \sum[R_{ref}]$ accounts for the reference-state subtraction. We diagonalize $H_{eff}(k)$ to obtain HF eigenvalues/eigenvectors, update $R(k)$ from the occupied HF states, and iterate until convergence ($\max_k |R_{new}(k) - R_{old}(k)| < 5 \times 10^{-6}$). In practice, we use linear mixing $R \leftarrow (1-\lambda)R_{old} + \lambda R_{new}$ with $\lambda = 0.2$ at each iteration, and the chemical potential $\mu$ is adjusted to enforce the filling $\nu$ under smearing temperature $k_BT_{HF}$.

To probe spontaneous symmetry breaking, we prepare the initial state as flavor-symmetric ground state of the non-interacting system with small random noise. The noise introduces a minor random flavor polarization and inter-valley coherence. We do not enforce time-reversal symmetry, so that converged solutions can be Chern insulators. After convergence, we compute the total Chern number $C$ of the occupied active bands by the Fukui–Hatsugai–Suzuki method on the **k**-mesh. We also track flavor polarization and inter-valley coherence by evaluating the spin/valley-resolved occupancies extracted from $R(k)$.

At $\nu = 1$, we find a fully spin-valley polarized HF ground state (see Extended Data Fig. 2a). At small displacement field (4 meV < $\Delta_U$ < 15 meV), the fully occupied active band has Chern number $C = \pm 4$ when polarized to the ± valley. Upon increasing $\Delta_U$ beyond a critical value $\Delta_{Uc} = 15.4$ meV, the occupied active band switches to $C = \pm 3$. By comparing with the previous section of continuum model, where the first band above CNP has the same Chern number in similar $\Delta_U$ regions, we can conclude that the topology of HF band inherits from the non-interacting band.

At $\nu = 2$, within the projected screened-Coulomb interaction, the HF energetics is symmetric in the flavor space. We therefore find a manifold of degenerate HF Slater determinants at $\nu = 2$: occupying active bands of any two among the four flavors yields the same HF energy within numerical accuracy, and the total Chern number $C = 0$ (±8) when the two flavors are in the opposite (same) valleys. Note that there is no spontaneous inter-valley coherence order throughout our HF calculations. However, in real devices, additional short-range intervalley exchange terms (Hund's or anti-Hund's coupling) lift this degeneracy. Importantly, irrespective of the sign of such intervalley coupling, the energy is lowered when both valleys are occupied, which selects a valley-unpolarized configuration and matches the experimentally observed trivial insulator ($C = 0$) at $\nu = 2$(see Fig. 1g).

At $\nu = 3$, the self-consistent solution (see Extended Data Fig. 2c) can be viewed as adding one extra spin-valley polarized flavor on top of the $\nu = 2$ background. The converged HF state remains polarized and yields a total Chern number $C = \pm 4$, consistent with experiment over the explored displacement-field range.

At fractional fillings around $\nu = 3$, the experiment reports a series of integer-Chern insulating states. As stated in the main text, for the $|C| > 4$ states, the non-monotonic temperature dependence of $\rho_{xy}$ is consistent with a scenario in which a less stable AHC on top of a $|C| = 4$ normal Chern-insulator background. We therefore view these states as suggestive, though far from conclusive, evidence for the long-sought AHC. Moreover, we conjecture that all the non-commensurate Chern insulators may share a similar mechanism, in which, on top of the background at the highest integer filling below $\nu$, spontaneous translation-symmetry breaking reconstructs the partially filled $|C| = 4$ band and opens spectral gaps with different occupied Chern numbers. However, a quantitative investigation of the mechanisms underlying these non-commensurate Chern insulators is beyond the scope of present work and is left for future study.

**FCI based on $C = 4$ parent band**

Now we explain how neutral particle-hole excitations can generate the required negative Hall conductance $\sigma_{xy}^{(3)} = -1/3$. We only consider one of the two fully occupied $C = 1$ bands in the positive valley and an empty $C = -1$ band in the negative valley. We label the holes in the $C = 1$ band and particles in the $C = -1$ band as $w_j$'s and $z_i$'s, respectively. Since both quasi-particles experience an effective negative magnetic field, we can write a Halperin state of the form

$$\Psi = \prod_{i<i'} (z_i^* - z_{i'}^*)^n \prod_{j<j'} \left(w_j^* - w_{j'}^*\right)^n \prod_{ij} \left(z_i^* - w_j^*\right)^m exp\left(-\frac{1-B}{4}\sum_i |z_i|^2 - \frac{1+B}{4}\sum_j \left|w_j\right|^2\right).$$

Here $B$ is the external magnetic field, which is chosen parallel to the $C = 1$ direction. Since $z_i$'s and $w_j$'s are from valleys with $C = -1$, 1, the corresponding flat-band degeneracies (effective Landau level degeneracy) decreases and increases with respect to $B$, respectively.

To calculate the fillings and Hall conductance, we notice that $|\psi|^2$ can be interpreted as the partition function of a classical Coulomb gas with the energy:

$$2n\, \nu_z^2 + 2n\, \nu_w^2 + 2m\, \nu_z \nu_w - 2(1-B)\nu_z - 2(1+B)\nu_w$$

Where $\nu_z$ and $\nu_w$ are the densities of particles and holes, respectively. Minimizing this energy gives

$$\begin{pmatrix} \nu_z \\ \nu_w \end{pmatrix} = \begin{pmatrix} n & m \\ m & n \end{pmatrix}^{-1} \begin{pmatrix} 1-B \\ 1+B \end{pmatrix} = \frac{1}{n^2 - m^2} \begin{pmatrix} n & -m \\ -m & n \end{pmatrix} \begin{pmatrix} 1-B \\ 1+B \end{pmatrix}$$

The physical charge density is then given by

$$\nu = \nu_z - \nu_w = -\frac{2}{n-m} B$$

The slope $-\frac{2}{n-m}$ should give the required negative Hall conductance -1/3. We hence choose $n = 7$, $m = 1$. In the absence of the external field $B$, the densities of both particles and holes equal $\frac{1}{n+m} = \frac{1}{8}$.

This state is characterized by the K matrix

$$K = -\begin{pmatrix} 7 & 1 \\ 1 & 7 \end{pmatrix}$$

with the t-vector $t = (1, -1)^T$, where the two components represent the electric charges for the particle and hole excitations.

Since both the $C = 1$ and $C = -1$ valleys host four layers, multiple Halperin-like states can, in principle, account for the total filling $\nu' = 8/3$ and Hall conductance $\sigma_{xy} = 7/3$. Distinguishing them is difficult from the current experimental data. Nevertheless, all such states are not fully valley-polarized and exhibit nontrivial braiding statistics between excitations in the two valleys.

## Method References

56. Feng, Z. *et al.* Rapid infrared imaging of rhombohedral graphene. *Phys. Rev. Appl.* **23**, 034012 (2025).
57. Zhang, H. *et al.* Moiré enhanced flat band in rhombohedral graphene. *Nat. Mater.* 1–7 (2025).

## Acknowledgements

We thank the Peking Nanofab for process support.

## Funding statement

X.L. discloses support for the research of this work from the National Key R&D Program (Grant No. 2022YFA1403500 and 2024YFA1409002), the National Natural Science Foundation of China (Grant No. 12521006 and 12274006) and Beijing National Laboratory for Condensed Matter Physics (Grant No. 2025BNLCMPKF001). Z.S. discloses support for the research of this work from National Natural Science Foundation of China (Grant No. 12274005), National Key R&D Program (Grant No. 2021YFA1401900) and Quantum Science and Technology-National Science and Technology Major Project (No. 2021ZD0302403). K.L. discloses support for the research of this work from National Natural Science Foundation of China (Grant No. 92577201, 12427806 and T2188101). W.W. discloses support for the research of this work from Quantum Science and Technology-National Science and Technology Major Project (Grant No. 2025ZD0300500) and the Peking University Boya Postdoctoral Fellowship. K.W. and T.T. discloses support for the research of this work from the JSPS KAKENHI (Grant nos. 21H05233 and 23H02052) and World Premier International Research Center Initiative (WPI), MEXT, Japan.

## Author Contributions

X.L., Z.S., K.L and W.W. supervised the project; Z. L. and W.W fabricated the devices with help from Q.Y.; Z. L. and W.W performed the measurement with help from Z.Z.; Z. L., W.W., F.W., X.X., J.W., X.L. analyzed the data; F.W. J.W. and Z.S. performed the theoretical modeling; T.T. and K.W. contributed materials; Z.L., W.W., F.W., J.W., K.L., Z.S. and X.L. wrote the paper.

## Competing interests

The authors declare no competing interests.

## Data Availability

All data supporting the findings of this study are available within the main text, figures and Supplementary Information, or from the corresponding authors upon request. Source data are provided with this paper.

# Extended Data Tables

| Device | Twist angle (°) | Moiré wavelength (nm) | FCI at ν = 2/3 with C = 7/3 | Displacement field modulated Chern insulators at ν = 1 | Cascades of Chern insulators around ν = 3 |
|---|---|---|---|---|---|
| D1 | 1.38 | 10.19 | Yes | C =3-4 | \|C\| = 2-7 |
| D2 | 1.37 | 10.29 | Yes | C =3-4 | \|C\| = 1-7 |
| D3 | 1.35 | 10.41 | No | C =3-4 | No |
| D4 | 1.28 | 11.04 | No | C =3-4 | No |
| D5 | 1.37 | 10.28 | Yes | C =3-4 | \|C\| = 2, 4-5 |

**Extended Data Table 1. | Device summary.** The twist angles, inferred from the carrier density at filling factor $\nu = 4$ are determined after gate capacitances are calibrated through Landau Fan diagrams. The moiré wavelength is calculated from the twist angles. Devices showing (not showing) fractional Chern insulators are noted as Yes (No).

# Extended Data Figures

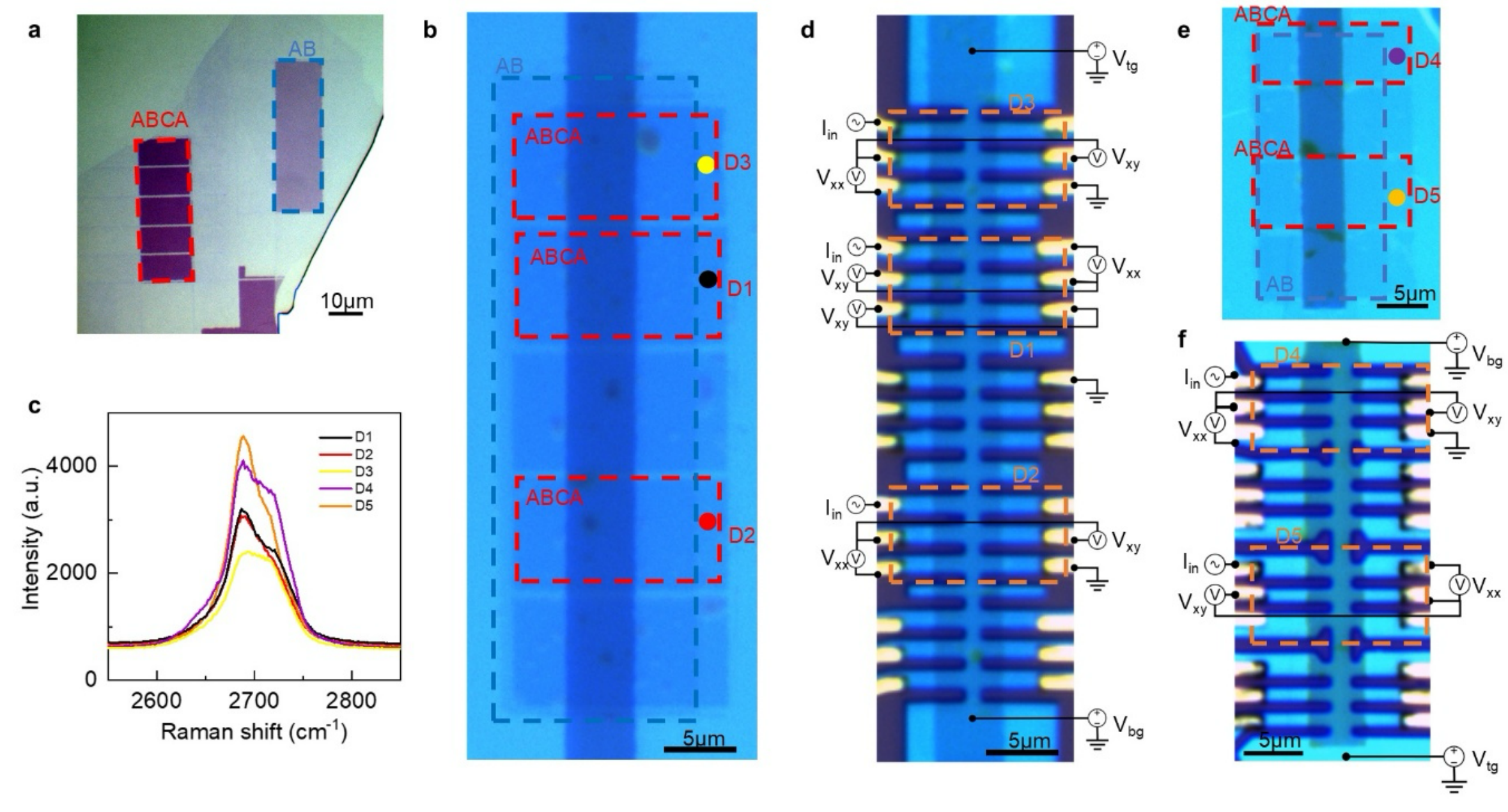

**Extended Data Fig. 1 | Structural characterization and optical micrographs of TBRTG devices.** **a**, Optical micrographs showing the bilayer graphene and rhombohedral tetralayer graphene regions after laser etching for device D1, D2 and D3. **b**, Optical image of the stacked devices D1, D2 and D3, where the blue and red dashed boxes indicate the bilayer and rhombohedral tetralayer graphene, respectively. **c**, Raman spectra acquired from the locations marked by the colored circles marked in **b** and **e**. **d**, Optical image of devices D1, D2 and D3 after nanofabrication with measurement setup. The regions corresponding to independent devices are outlined by orange dashed squares. **e,** Optical image of the stacked devices D4 and D5. **f,** Optical image of devices D4 and D5 after nanofabrication with measurement setup.

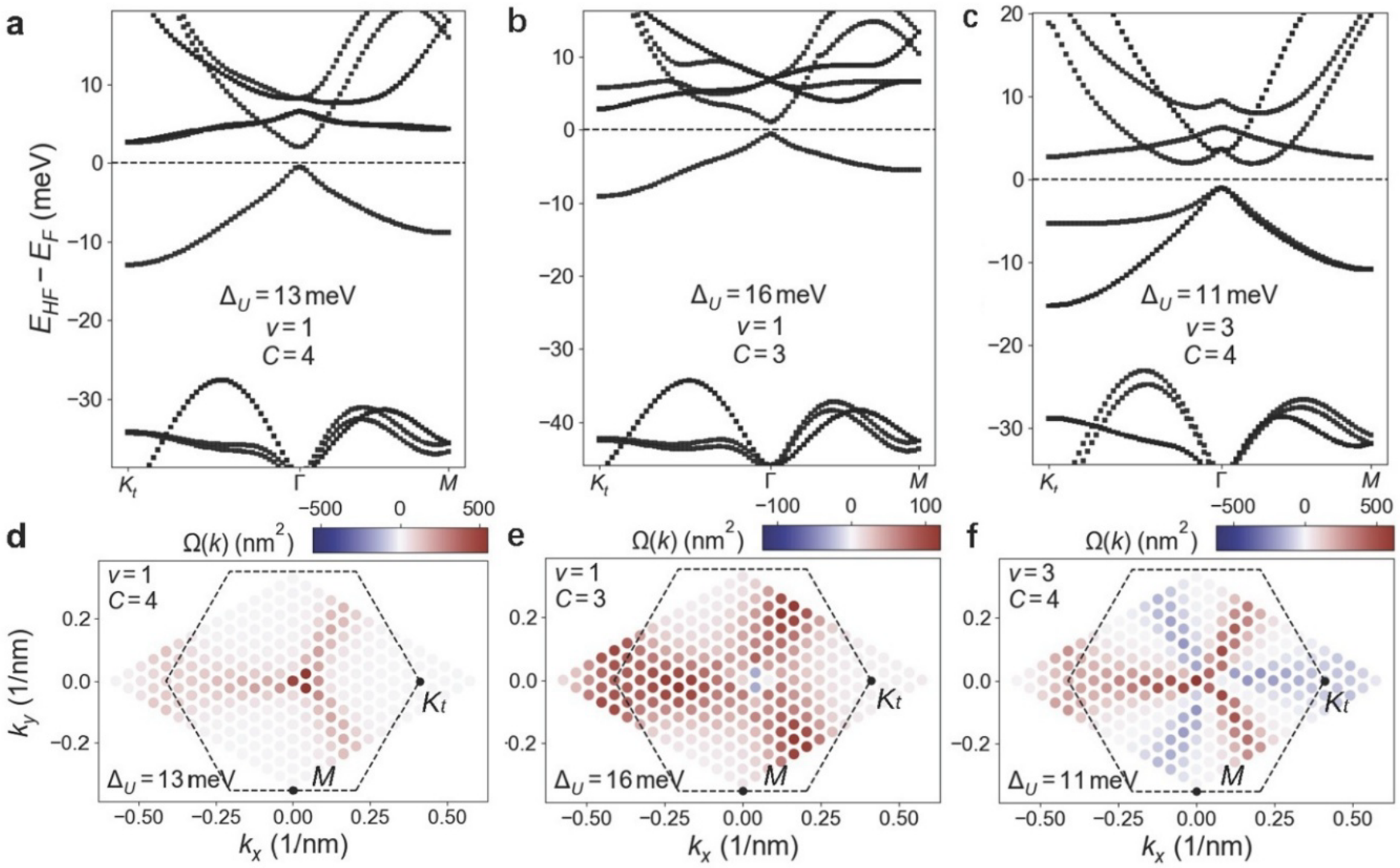

**Extended Data Fig. 2 | Hartree-Fock calculation results. a-c,** Self-consistent HF band eigenvalues along the high symmetry line $K_t - \Gamma - M$ in the moiré Brillouin zone. The HF self-consistency is reached on a 15 × 15 **k**-mesh while band eigenvalues are evaluated and plotted on a finer **k**-mesh. For all the HF calculations, $U_\xi = 24$ meV and $k_B T_{HF} = 0.0026$ meV. The dashed horizontal line indicates the Fermi energy. All the plotted bands are included in the HF self-consistency. $\Delta_U$ is the potential difference between adjacent graphene layers. **(a)** $\nu = 1$ $\Delta_U = 13$ meV: the occupied active HF band is polarized to the flavor (valley $\eta$, spin $s$) = (+, ↑) and has Chern number $C = 4$. **(b)** $\nu = 1$ $\Delta_U = 16$ meV: the occupied active HF band is (+, ↑)-polarized with $C = 3$. **(c)** $\nu = 3$ $\Delta_U = 11$ meV: the 3 occupied active HF bands are partially polarized toward (+, ↑), which can be decomposed into an unpolarized two-band background plus one (+, ↑)-polarized band, yielding $C = 4$. **d-f,** Total Berry curvature $\Omega(k)$ summed over the occupied active HF bands, and the data are extracted from the same HF calculations of **(a-c)**, respectively**.** Note that, depending on the initial conditions, the HF iteration can converge to other spin/valley-polarized solutions with identical energies, indicating spontaneous breaking of time-reversal symmetry and spin SU(2) symmetry.

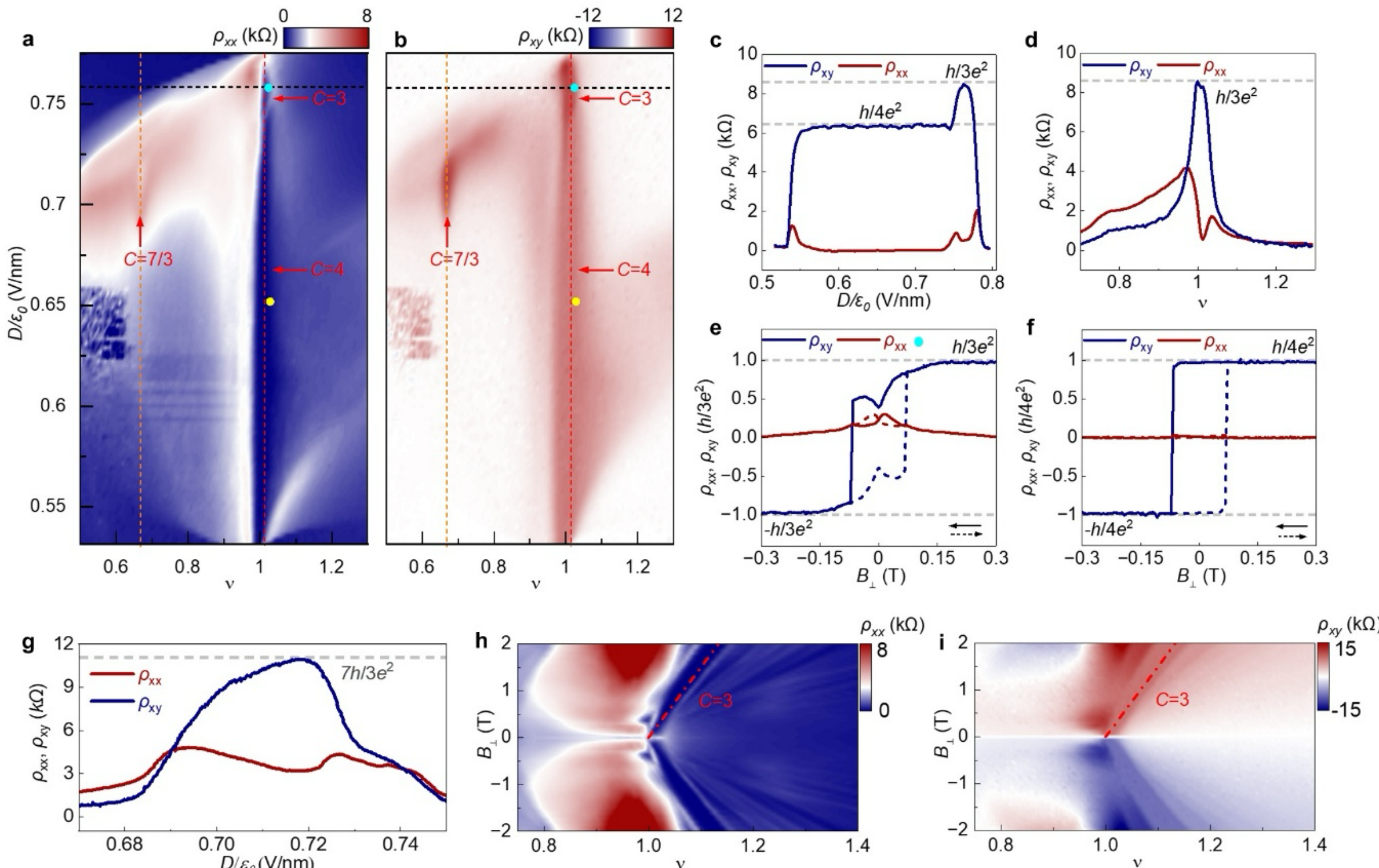

**Extended Data Fig. 3 | Displacement field modulated Chern insulators at $\nu = 1$ for device D2.** **a-b,** Symmetrized $\rho_{xx}$ **(a)** and anti-symmetrized $\rho_{xy}$ **(b)** as functions of $\nu$ and $D/\varepsilon_0$ measured at perpendicular magnetic field $B_\perp = \pm 0.1$ T and temperature $T = 10$ mK around $\nu = 1$. The red dashed line is the linecut for **c**. The black dashed line is the linecut for **d**, **h** and **i**. The orange dashed line is the linecut for **g**. The cyan and yellow circles represent the positions for **e** and **f** respectively. **c,** Symmetrized $\rho_{xx}$ and anti-symmetrized $\rho_{xy}$ as functions of $D/\varepsilon_0$ at $\nu = 1.01$ and $B_\perp = \pm 0.3$ T. **d,** Symmetrized $\rho_{xx}$ and anti-symmetrized $\rho_{xy}$ as functions of $\nu$ at and 0.756 V/nm. **e-f,** Hysteresis loops of Symmetrized $\rho_{xx}$ and anti-symmetrized $\rho_{xy}$ as functions of $B_\perp$ at $\nu = 1.02$, $D/\varepsilon_0 = 0.756$ V/nm **(e)** and $\nu = 1.02$, $D/\varepsilon_0 = 0.652$ V/nm **(f)**. **g,** Symmetrized $\rho_{xx}$ and anti-symmetrized $\rho_{xy}$ as functions of $D/\varepsilon_0$ at $\nu = 2/3$. **h-i,** Symmetrized $\rho_{xx}$ and anti-symmetrized $\rho_{xy}$ as functions of $\nu$ and $B_\perp$ at $D/\varepsilon_0 = 0.756$ V/nm. The red dashed lines indicate the $\rho_{xx}$ minimal and $\rho_{xy}$ plateaus, which strictly follow the Středa formula $C = h/e(\partial n/\partial B)$.

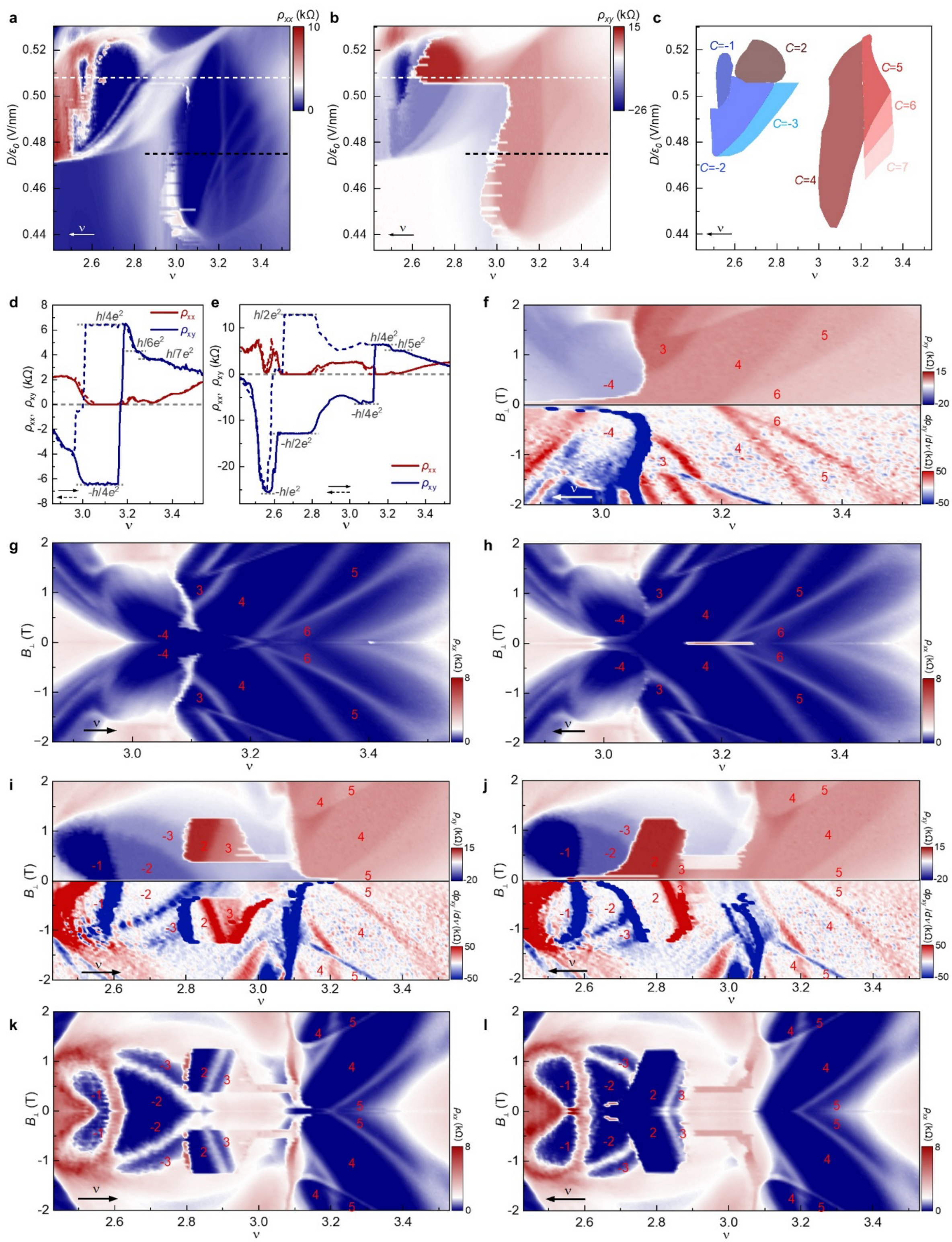

**Extended Data Fig. 4 | Cascades of high Chern insulators around ν = 3 and the hysteretic behavior for device D2. a-b,** Symmetrized $\rho_{xx}$ **(a)** and anti-symmetrized $\rho_{xy}$ **(b)** as functions of $\nu$ and $D/\varepsilon_0$ measured at perpendicular magnetic field $B_\perp = \pm 0.1$ T and temperature $T = 10$mK with negative $\nu$ sweeping direction for device D1. The black and white dashed lines represent the linecuts for **d** and **e** respectively. **c,** Schematic of the phase diagram of **a-b**, where the Chern

insulators with different Chern numbers are highlighted. **d-e,** Symmetrized $\rho_{xx}$ and anti-symmetrized $\rho_{xy}$ as functions of $\nu$ at $B_\perp = \pm 0.1$ T for $D/\varepsilon_0 = 0.475$ V/nm **(d)** and $D/\varepsilon_0 = 0.508$ V/nm **(e)** with $\nu$ sweeping positively (solid lines) and negatively (dashed lines). **f,** Anti-symmetrized $\rho_{xy}$ and derivative of anti-symmetrized $\rho_{xy}$ on $\nu$ as functions of $\nu$ and $B_\perp$ at $D/\varepsilon_0 = 0.475$ V/nm with negative $\nu$ sweeping direction. **g-h,** Symmetrized $\rho_{xx}$ as functions of $\nu$ and $B_\perp$ at $D/\varepsilon_0 = 0.475$ V/nm with positive **(i)** and negative **(j)** $\nu$ sweeping direction. **i-j,** Anti-symmetrized $\rho_{xy}$ and derivative of anti-symmetrized $\rho_{xy}$ on $\nu$ as functions of $\nu$ and $B_\perp$ at $D/\varepsilon_0 = 0.508$ V/nm with positive **(i)** and negative **(j)** $\nu$ sweeping direction. **k-l,** Symmetrized $\rho_{xx}$ as functions of $\nu$ and $B_\perp$ at $D/\varepsilon_0 = 0.508$ V/nm with positive **(k)** and negative **(l)** $\nu$ sweeping direction.

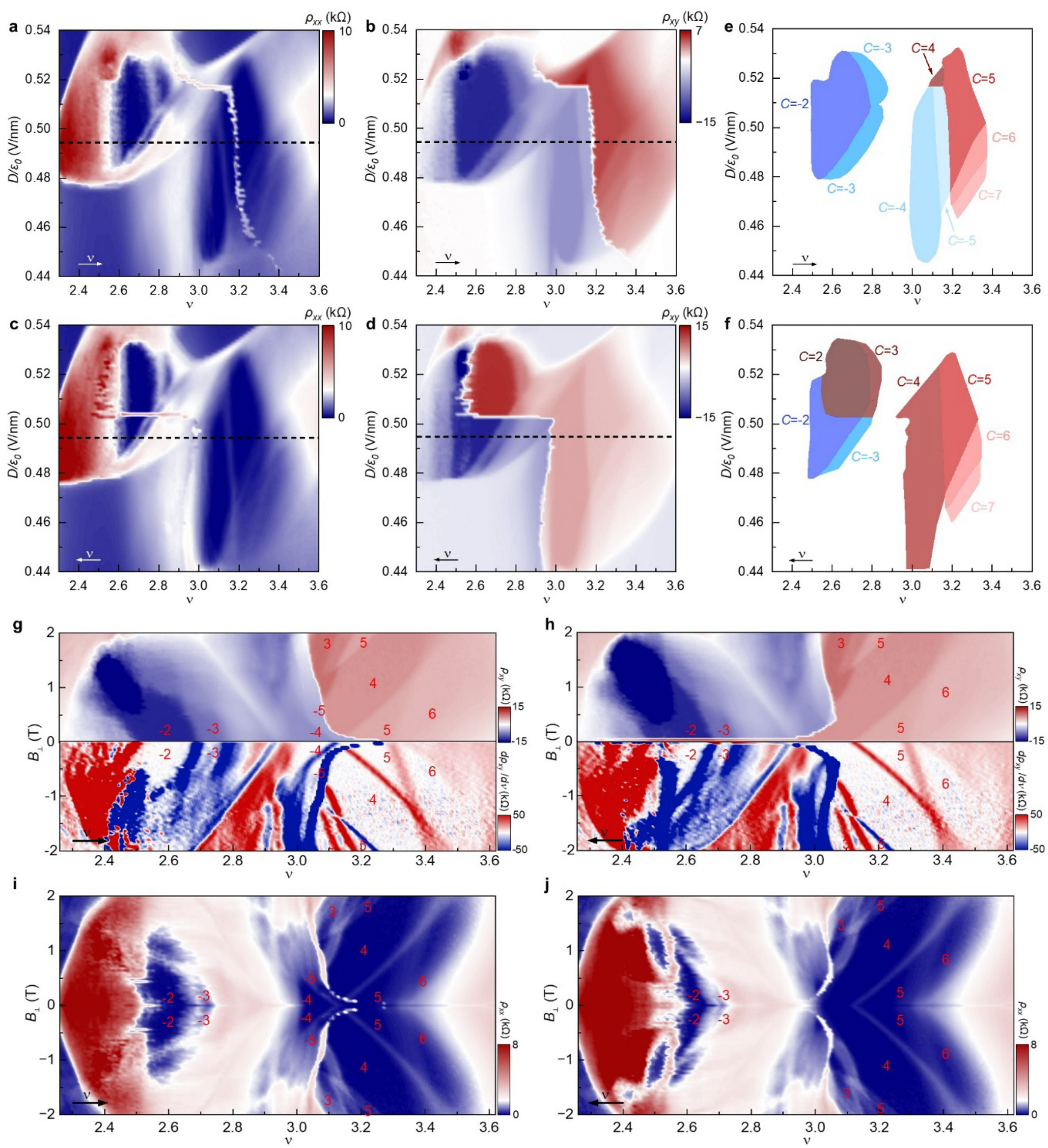

**Extended Data Fig. 5 | Phase diagram and complex fan diagram of high Chern insulators around $\nu$ = 3 for device D1. a-d,** Symmetrized $\rho_{\mathrm{xx}}$ **(a, c)** and anti-symmetrized $\rho_{\mathrm{xy}}$ **(b, d)** as functions of $\nu$ and $D/\varepsilon_0$ measured at perpendicular magnetic field $B_\perp = \pm 0.1$ T and temperature $T$ = 10 mK with positive **(a-b)** and negative **(c-d)** $\nu$ sweeping direction for device D2. The black dashed lines represent the linecuts for **g** and **h**. **e-f,** Schematics of the phase diagram of **a-b (e)** and **c-d (f)**, where the Chern insulators with different Chern numbers are highlighted. **g-h,** Anti-symmetrized $\rho_{\mathrm{xy}}$ and derivative of anti-symmetrized $\rho_{\mathrm{xy}}$ on $\nu$ as functions of $\nu$ and $B_\perp$ at $D/\varepsilon_0$ = 0.508 V/nm with positive **(g)** and negative **(h)** $\nu$ sweeping direction. **i-j,** Symmetrized $\rho_{\mathrm{xx}}$ as a function of $\nu$ and $B_\perp$ at $D/\varepsilon_0$ = 0.508 V/nm with positive **(i)** and negative **(j)** $\nu$ sweeping direction.

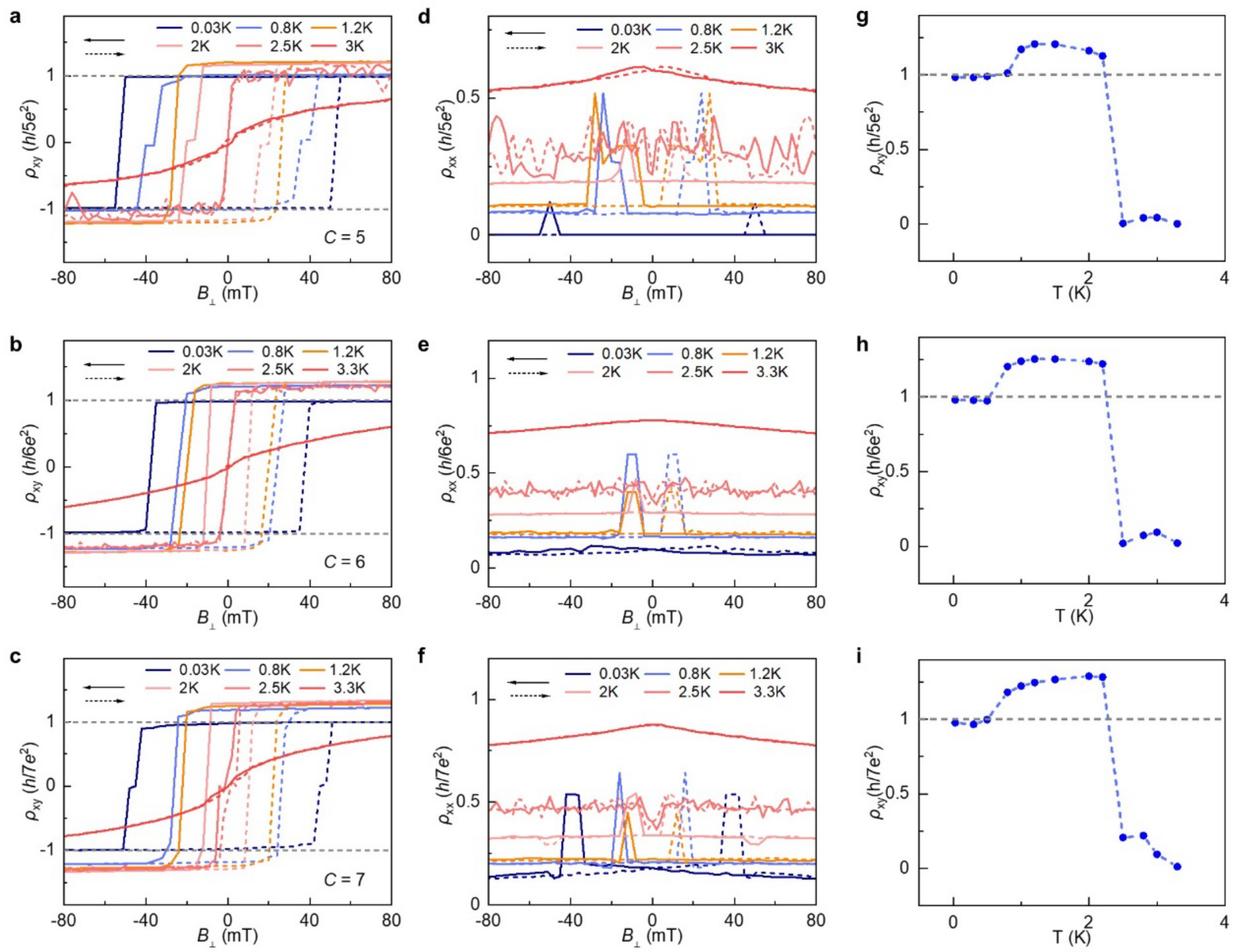

**Extended Data Fig. 6 | Magnetic hysteresis and temperature evolution of Hall resistance near $\nu \approx 3$ at selected displacement fields for device D1. a-c,** Anti-symmetrized $\rho_{\mathrm{xy}}$ hysteresis loops measured at various temperatures for $\nu$ = 3.23 at $D/\varepsilon_0$ = 0.503 V/nm, $\nu$ = 3.29 at $D/\varepsilon_0$ = 0.484 V/nm, and $\nu$ = 3.29 at $D/\varepsilon_0$ = 0.474 V/nm. **d-f,** Corresponding Symmetrized $\rho_{\mathrm{xx}}$ hysteresis loops acquired under the same conditions. **g-i,** Temperature dependence of $\rho_{\mathrm{xy}}$ at $B_\perp$ = 0 T ,which shows a pronounced saturation trend as the temperature decreases.

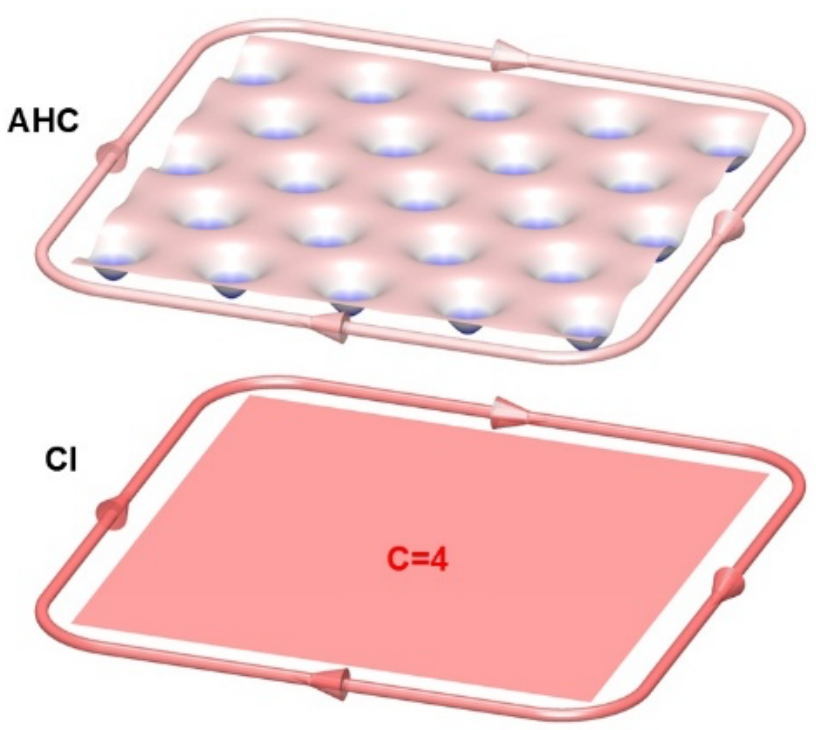

**Extended Data Fig. 7 | Conceptual illustration of cascades of Chern insulators around $\nu = 3$.** A partially filled electron liquid ($\nu' = \nu - 3$) with broken translation symmetry and quantized Berry curvature (Anomalous Hall crystal) is on top of a robust $|C| = 4$ Chern insulator background.

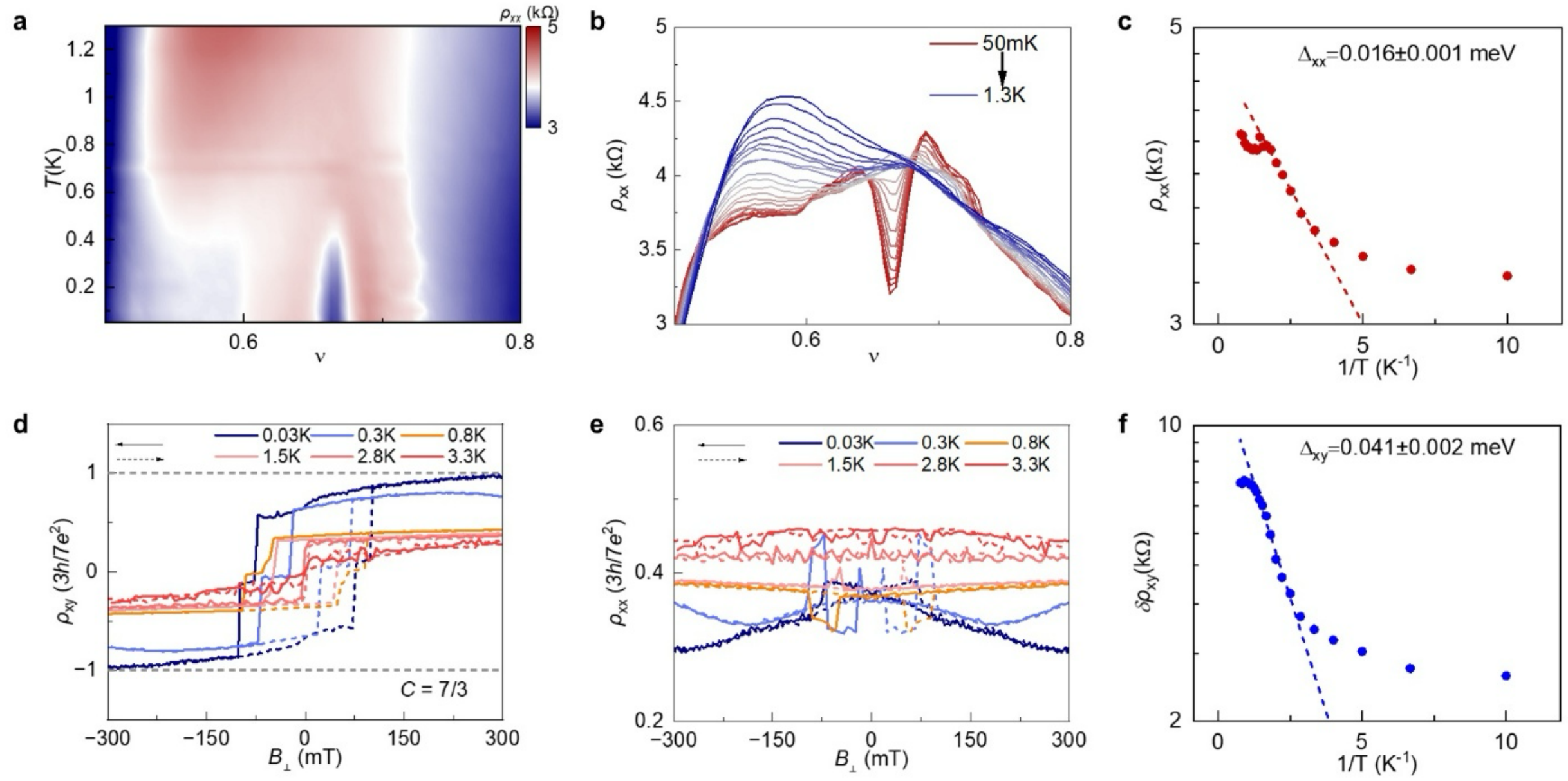

**Extended Data Fig. 8 | Temperature dependence and the thermal activation gap of the FCI state for device D2**. **a,** Colormap of symmetrized $\rho_{xx}$ (measured at $B_\perp = \pm 0.1$ T, $D/\varepsilon_0 = 0.714$ V/nm). **b,** Linecut traces of $\rho_{xx}$ versus $T$ from **a**. **c**, Thermal activation gap ($\Delta_{xx}$) extracted by fitting symmetrized $\rho_{xx}$ versus $1/T$ at $B_\perp = 0.1$ T. **d-e,** Magnetic hysteresis loops of anti-symmetrized $\rho_{xy}$ (**d**) and symmetrized $\rho_{xx}$ (**e**) measured at different temperatures for $\nu = 2/3$ at $D/\varepsilon_0 = 0.713$ V/nm. **f,** Thermal activation gap ($\Delta_{xy}$) extracted by fitting anti-symmetrized $\delta\rho_{xy} = h/(Ce^2) - \rho_{xy}$ versus $1/T$ at $B_\perp = 0.1$ T.

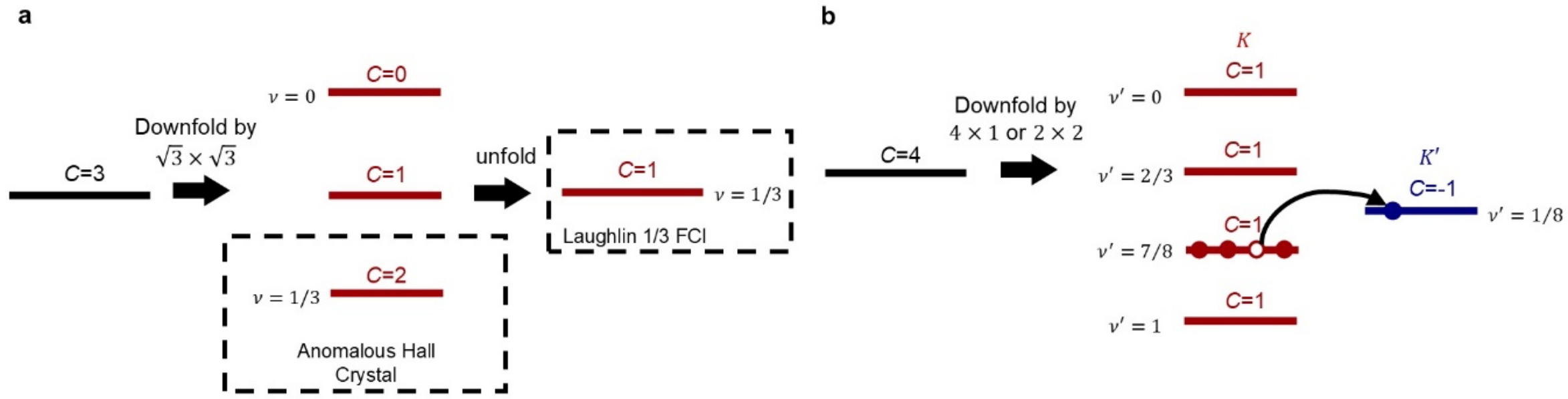

**Extended Data Fig. 9 | possible mechanisms of the formation of the *C* = 7/3 fractional Chern insulator at *ν* = 2/3. a,** The $C = 3$ parent band is down folded into a Chern structure (0, 1, 2) or (1, 0, 2). The lowest folded band with $C = 2$ is filled by $\nu^{(1)} = 1/3$ and contributes a Hall conductivity $\sigma_{xy}^{(1)} = 2e^2/h$. The remained electrons with filling $\nu^{(2)} = 1/3$ occupies the $C = 1$ band and fractionalize into a Laughlin $C = 1/3$ fractional Chern insulator and contributes a Hall conductivity $\sigma_{xy}^{(1)} = (1/3)e^2/h$. **b,** The $C = 4$ parent band is mapped to four-layer $C = 1$ bands. Two $C = 1$ bands are fully occupied by $\nu'^{(1)} = 2$, contributing to $\sigma_{\mathrm{xy}}^{(1)} = 2e^2/h$. Another $C = 1$ bands is occupied by $\nu'^{(2)} = 2/3$, contributing to $\sigma_{\mathrm{xy}}^{(2)} = (2/3)e^2/h$. The remaining $\sigma_{\mathrm{xy}}^{(3)} = (-1/3)e^2/h$ is obtained by electron- hole excitation distributed over the two valley, which is described by a Halperin wavefunction.